\documentclass[preprint,showpacs,preprintnumbers,amsmath,amssymb,prb]{revtex4}
\usepackage{graphicx}
\usepackage{dcolumn}
\usepackage{bm}
\usepackage{epsfig}
\begin{document}

\preprint{PREPRINT}

\title{Core-softened Fluids,  Water-like Anomalies and  the Liquid-Liquid Critical Points}

\author{Evy Salcedo}
\affiliation{Departamento de F\'{\i}sica,
Universidade Federal de Santa Catarina, Florian\'opolis, SC, 
88010-970, Brazil}
\email{esalcedo@fsc.ufsc.br}

\author{ Alan Barros de Oliveira}
\affiliation{Departamento de F\'{i}sica, Universidade Federal de Ouro Preto,
Ouro Preto, MG, 35400-000, Brazil} 
\email{oliveira@iceb.ufop.br}

\author{Ney M. Barraz Jr.}
\affiliation{Programa de P\'os-gradua\c c\~ao em F\'{\i}sica 
da UFRGS, Bento Gon\c calves 9500, 91501970, Porto Alegre, RS, Brazil} 
\email{neybarraz@gmail.com}

\author{Charusita Chakravarty}
\affiliation{Department of Chemistry,
Indian Institute of Technology-Delhi,
New Delhi, 110016, India}
\email{charus@chemistry.iitd.ernet.in }

\author{ Marcia C. Barbosa}
\affiliation{ Instituto de F\'{\i}sica,Universidade Federal do Rio Grande do Sul,  Porto Alegre, RS,  1501-970, Brazil}
\email{marcia.barbosa@ufrgs.br}
\begin{abstract}

Molecular dynamics simulations are used to  examine the relationship 
between water-like anomalies and the liquid-liquid critical point in a 
family of model fluids with multi-Gaussian,   core-softened pair interactions. 
The core-softened pair interactions have two length scales, such
that the longer length scale associated with
a shallow, attractive well is kept constant while the shorter length scale
associated with the repulsive shoulder is varied  from an inflexion point to
a minimum of progressively increasing depth. The maximum depth of the shoulder
well is chosen so that the resulting potential reproduces the
oxygen-oxygen radial distribution function of the ST4 model of water.
  As the shoulder well depth increases,
the pressure required to form the high density liquid decreases and the
temperature up to which the high-density liquid is stable increases, resulting in
the shift of the liquid-liquid critical point to much lower pressures and
higher temperatures. To understand the entropic effects associated with the
changes in the interaction potential,  the pair correlation 
entropy is computed  to show that the excess entropy anomaly diminishes when
the shoulder well depth increases. Excess entropy
scaling of diffusivity in this class of fluids is demonstrated, showing that
decreasing strength of the excess entropy anomaly with increasing shoulder depth 
results in  the progressive loss of water-like thermodynamic, structural and transport anomalies.
Instantaneous normal mode analysis was used to index the overall
curvature distribution of the fluid and the  fraction of imaginary frequency modes 
was shown to correlate well with the anomalous behaviour of the diffusivity and 
the pair correlation entropy. The results suggest in the case of core-softened potentials, 
in addition to the presence of  two length scales, 
energetic and entropic effects associated with local
minima and curvatures of the pair interaction play an important
role in determining the presence of water-like anomalies and the liquid-liquid
phase transition. 
\end{abstract}


\maketitle

\section{Introduction}

Water is characterized by well-known thermodynamic and kinetic liquid-state anomalies;
for example, the rise in density on isobaric heating (density anomaly) and 
the increase in molecular mobility on isothermal compression (diffusivity anomaly)
\cite{ms98,pgd03}.
Since the anomalies of bulk water are  connected with its behaviour as
a solvent in chemical and biological systems, an understanding of the structural
origins of such anomalous behaviour has attracted considerable attention\cite{dtvh05,kfs08}.
While the anomalies of water were initially presumed to be uniquely connected
to the hydrogen-bonded network of water\cite{bf33}, there is now evidence that
a  number of liquids display water-like liquid-state anomalies,  such as Te~\cite{Th76} Ga, Bi~\cite{LosAlamos}, S~\cite{Sa67,Ke83} Ge$_{15}$Te$_{85}$~\cite{Ts91},
silica~\cite{An00,Ru06b,Sh02,Po97}, silicon~\cite{Ta02} and BeF$_2$~\cite{An00,scc06,asc07,ac07,ac09pre,agc09}. The generic relationships between structure, entropy and mobility underlying
this diverse set of liquids with water-like anomalies, can be understood in terms of the behaviour of the excess entropy ($S_{ex}$), defined as the difference between the entropy ($S$) of the
liquid and the corresponding ideal gas at the same density and temperature~\cite{scc06,assac10,Ol06a,oscb10,met061,met062,etm06,ybs08}. A necessary condition for the fluid to show
water-like thermodynamic and transport anomalies is an excess entropy anomaly, corresponding to a rise in excess entropy on isothermal compression. The structural basis for the
excess entropy anomaly is the existence of distinct forms of local order or length scales in the low- and high-density regimes; competition between the two types of local order results in a rise in excess entropy at intermediate densities.

In addition to the singularity-free scenario for water-like thermodynamic and kinetic anomalies, it has been conjectured that the anomalies of water are due to the presence of a
second liquid-liquid critical point, corresponding to the onset of a line of  first-order phase transitions between high- and low-density phases of water
\cite{pses92}. The relationship between the liquid-liquid critical point and water-like anomalies can be addressed by considering  minimal models of liquids with isotropic, pair-additive
interactions that  give rise to water-like anomalies, as well as liquid-liquid critical points~\cite{fmsbs01,ssbs98}.  Such models demonstrate that the presence of two repulsive
length scales in the pair interaction is necessary in order to give rise to liquid-state anomalies. If, in addition to two repulsive length scales, the pair interaction has an attractive
component, the fluid can show a liquid-liquid critical point (LLCP), in addition to the liquid-gas critical point. While two length scales in the pair interaction appears to be a necessary
condition for seeing both the LLCP and water-like anomalies, it is possible to design  isotropic  potentials with two length scales where appropriate variation of parameters can result in
shifting either  the LLCP or the water-like anomalies into the metastable or unstable regime~\cite{Ba09,Ol06a}.

In this paper, we study a family of liquids with continuous, core-softened pair interactions consisting of a hard core, a short-range shoulder and an attractive well at a larger
separation~\cite{Ba09}. Since the potentials share a common functional form consisting of a sum of one Lennard-Jones and four Gaussian terms, we refer to them as the family of
multi-Gaussian water-like liquids. By suitably varying the parameters, the outer attractive well can be left unchanged, while the shoulder can be progressively altered from being purely
repulsive to a deep attractive well. As the shoulder shifts from being purely repulsive to more attractive,   the  anomalous regime in the pressure-temperature plane shrinks and disappears
while the LLCP shifts to higher temperatures and lower pressures. The connection with atomistic models is made by ensuring that   the limiting case of the double minimum potential that has
no anomalies  corresponds to an isotropic potential that reproduces the oxygen-oxygen radial distribution function of ST4 water~\cite{hs93}. Using this set of anomalous fluids, we  address a
number of questions related to the features of the pair interaction, in addition to the two length scales, that control the temperature-pressure regime of the water-like liquid state
anomalies versus the liquid-liquid critical point. Both the liquid-liquid critical point and the water-like anomalies require a  change in the nature of local order in the liquid with
density, and therefore two length scales in the case of core-softened fluids. The liquid-liquid critical point, however, depends on the energetic bias towards segregation of the two length
scales with decreasing temperature. In contrast, the water-like liquid state anomalies  require an excess entropy anomaly, involving a  continuous  transformation of
the liquid  from low- to high-density through a range of quasi-binary states reflecting the competition between two length scales in the intermediate regime~\cite{Ol06a}.
In order to understand the relationship between the interaction potential, the water-like liquid state anomalies and the liquid-liquid critical point, it is therefore
necessary to consider the temperature-dependent stabilization of the low- and high-density length scales as well as the density-dependent changes in the entropy of the
system. 

In order to understand how the change in interaction potential within the multi-Gaussian water models affects the thermodynamic and kinetic water-like anomalies, it is necessary to map out
the excess entropy anomaly  for the different model fluids. We use the pair correlation entropy as a simple structural estimator of the excess entropy, defining it for a one-component
fluid of structureless particles as:
\begin{equation}
\frac{S_2}{Nk_B } = -2\pi\rho \int_0^\infty 
\{g(r)\ln g(r) -\left[ g(r)-1\right] \} r^2 dr 
\label{s2}
\end{equation}
where $g(r)$ is the radial distribution function. It is typically the dominant contribution to the excess entropy of a fluid expressed as a multi-particle correlation expansion of the form:
\begin{equation}
S_{ex}=S-S_{id}=S_2 +S_3 + \dots
\end{equation}
where $S_n$ is the entropy contribution due to $n$-particle spatial correlations~\cite{hsg52,ng58,hjr71,dcw87,be89}. The excess entropy and mobility anomalies are linked by excess entropy
scaling  relations of the form
\begin{equation} 
X^*=A\exp (\alpha (S_{ex}/Nk_B))\; ,
\end{equation}
where $X^*$ are dimensionless transport properties with either macroscopic
(Rosenfeld) or microscopic (Dzugutov) reduction
parameters and the scaling parameters, $\alpha$ and $A$, depend on the functional form 
of the underlying interactions~\cite{yr771,yr772,yr99,md96}.

As an additional means to relate the interaction potential to the liquid-state
properties, we characterize the potential energy surface (PES)  of the multi-Gaussian
family of water-like liquids using instantaneous normal mode analysis. In the   Instantaneous Normal Mode (INM)  approach, the key quantity
is the  ensemble-averaged curvature distribution of the PES sampled by the  system.  For a system of $N$ particles, the mass-weighted Hessian associated with 
each instantaneous configuration is diagonalized to yield $3N$ normal mode eigenvalues and eigenvectors and the ensemble-average of this
distribution is referred to as the INM spectrum.  The INM spectrum of a liquid will have a substantial fraction of unstable modes with 
negative eigenvalues that simulations suggest is strongly correlated with the  diffusivity~\cite{sc01a,sc021,sc022,cfsos94,lssss00,rm97}. Random energy models of liquids also suggest
that for supercooled liquids there will be a connection between the fraction of imaginary modes, the diffusivity and the configurational entropy~\cite{tk00,kck02}. Our  previous work on
INM analysis of a  core-softened water-like fluid demonstrated that the instantaneous normal mode spectra carry significant information on the dynamical consequences of the interplay between
length scales characteristic of anomalous fluids~\cite{oscb10}. We have therefore performed an INM analysis of the multi-Gaussian water-like fluids to understand the relationship between
the interaction potential, anomalies and the liquid-liquid critical point.

The paper is organized as follows. The computational details of the simulations and the equation of state data for the multi-Gaussian family of water models are summarized in Section II.
Section III contains the results and the conclusions in Section IV.

\section{The model} 
\label{sec:model}

\subsection{The potential} 
\label{sec:potential}
The multi-Gaussian family of water-like fluids is defined by pair-additive,
continuous, core-softened interactions with the functional form
\begin{equation}
  U(r) = \epsilon \left[ \left( \frac{\sigma}{r} \right)^{a} -
    \left( \frac{\sigma}{r} \right)^{b} \right] + \sum_{j=1}^{4}h_{j}
  \exp \left[ -\left( \frac{r-c_{j}}{w_{j}} \right)^{2} \right] \;\;.
  \label{eq:potential}
\end{equation}

\begin{figure}[h]
  \begin{centering}
    \includegraphics[clip=true,width=8cm]{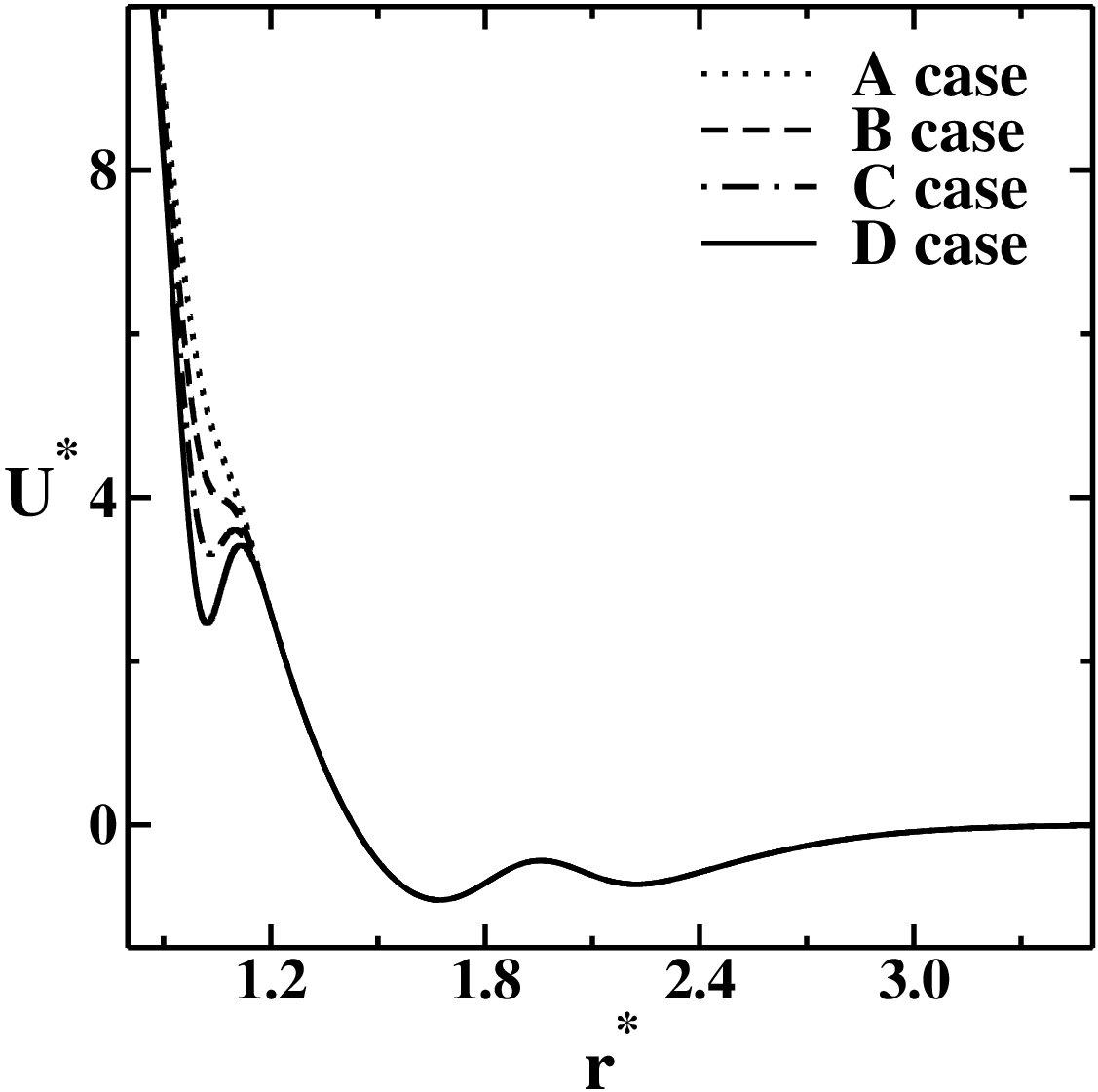} 
    \par
  \end{centering}
  \caption{Interaction potential obtained by changing parameters $h_1$ in Eq.~(\ref{eq:potential}). The potential and the distances are  in dimensionless
	    units $U^*=U/\gamma$ and $r^*=r/r_0$.}\label{fig:potential}
\end{figure}

The first term is a Lennard-Jones potential-like  and the second one is composed of four Gaussians, each of width $w_j$ centered at $c_j$. 
The potential and the distances are given in dimensionless units, $U^*=U/\gamma$ and $r^*=r/r_0$ where $\gamma$ is the  energy scale and
$r_0$ is the length scale chosen so the closest approach between particles is about $r^*=1$, i.e., so that the second length scale associated with the
repulsive shoulder remains the same.   Here we use $\epsilon/\gamma=0.02$ and $\sigma/r_0=1.47$. Modifying $h_1$ in the Eq.~(\ref{eq:potential}) allows us to change the depth 
of the hard-core well, as illustrated in Fig.~\ref{fig:potential} while keeping the shape  and location of the attractive well constant. We report here results for  four different values for
$h_1$ and they are expressed as a multiple of a reference  value $h_1^{ref}$ as shown in the Table~\ref{table1}. For all the four cases  the values of 
$a,b,\{c_j, w_j\}$ with $j=1, \dots, 4$ and $h^{ref}$. Table~\ref{table2}  gives the parameter values in $\AA$ and $kcal/mol$ 
consistent with reproducing the oxygen-oxygen radial distribution of ST4 water using case D~\cite{HG93}.

\begin{center}
\begin{table}
\caption{Parameters $h_1$ for potentials A, B, C and D.}
\centering{}
\begin{tabular}{|c|c|}
\hline 
Potential &   Value   of $h_1$  \tabularnewline \hline \hline
$A$       & $ 0.25\, h_1^{ref}$  \tabularnewline \hline
$B$       & $ 0.50\, h_1^{ref}$ \tabularnewline \hline
$C$       & $ 0.75\, h_1^{ref}$ \tabularnewline \hline
$D$       & $ 1.00\, h_1^{ref}$        \tabularnewline \hline

\end{tabular}
\label{table1}
\end{table}
\end{center}
\begin{center}
\begin{table}
\caption{Parameters for potentials A, B, C and D in units
of $\AA$ and of $kcal/mol$.}
\centering{}
\begin{tabular}{|c|c|c|c|}
\hline 
Parameter &    Value     & Parameter & Value                       \tabularnewline \hline \hline
$a$       & \ $9.056$ \  & $w_{1}$   & \hspace{0.2cm} $0.253$      \tabularnewline \hline
$b$       & $4.044$      & $w_{2}$   & \hspace{0.2cm} $1.767$      \tabularnewline \hline
$\epsilon$       & $0.006$      & $w_{3}$   & \hspace{0.2cm} $2.363$      \tabularnewline \hline
$\sigma$       & $4.218$      & $w_{4}$   & \hspace{0.2cm} $0.614$      \tabularnewline \hline
$c_{1}$   & $2.849$      & $h_{1}^{ref}$ & $-1.137$                \tabularnewline \hline
$c_{2}$   & $1.514$      & $h_{2}$   & \hspace{0.2cm} $3.626$      \tabularnewline \hline
$c_{3}$   & $4.569$      & $h_{3}$   &  $-0.451$                   \tabularnewline  \hline
$c_{4}$   & $5.518$      & $h_{4}$   & \hspace{0.2cm} $0.230$      \tabularnewline \hline
\end{tabular}
\label{table2}
\end{table}
\end{center}

\subsection{The simulation details} 
\label{sec:simulation}

The properties of the system were obtained by $NVT$ molecular 
dynamics using Nose-Hoover 
heat-bath with coupling parameter $Q = 2$. The system  is characterized by 
500 particles 
in a cubic box with periodic boundary conditions, 
interacting with the intermolecular potential described above. All physical 
quantities 
are expressed in reduced units and defined as
\begin{eqnarray}
t^* &=& \frac{t(m/\gamma)^{1/2}}{r_0}\nonumber  \\
T^* &=& \frac{k_{B}T}{\gamma }\nonumber  \\
p^*& =& \frac{pr_0}{\gamma }\nonumber \\
\rho^{*} &=& \rho r_0^3 \nonumber \\
D^* &=& \frac{D m}{\gamma r_0^2} \;\;.
\end{eqnarray}

Standard periodic boundary conditions together with 
predictor-corrector 
algorithm were used to integrate the equations of motion with a 
time step $\Delta t^{*}=0.002$ and potential cut off radius $r_{c}^{*}=3.5$.
The initial configuration is set on solid or liquid state and, in both cases, the
 equilibrium state was reached after $t_{eq}^{*}=1000$
(what is in fact $500 000$ steps since $\Delta t^{*}=0.002$) . From this time 
on the physical 
quantities were stored in intervals of $\Delta t_R^* = 1$ during
$t_R^* = 1000$. The system is uncorrelated after $t_d^*= 10$, as judged from the velocity 
auto-correlation function. $50$
descorrelated samples were used to get the average of the physical quantities. 

At each state point, 100 configurations were sampled and used to
construct the instantaneous normal mode spectra and associated quantities.
We repeated the calculation for some state points using 500 configurations 
and found no significant difference.

\subsection{Instantaneous Normal Modes Analysis}

The 
potential energy of  configuration ${\bf r}$ near ${\bf r_0}$ 
can be written as a Taylor expansion of the form:
\begin{eqnarray}
U({\bf r})=U({\bf r_0})-{\bf F} \bullet {\bf z}+\frac{1}{2} {\bf r^T} 
\bullet {\bf H} \bullet {\bf z}
\label{eq:expansion}
\end{eqnarray}
where ${\bf z_i}=\sqrt{m_i}({\bf r_i}-{\bf r_0})$ are the mass-scaled position 
coordinates of a particle $i$.
The first and second derivatives of $U({\bf r})$ with respect to the vector 
${\bf z}$  are the force and the Hessian matrix, denoted by 
${\bf F}$ and ${\bf H}$ respectively.
The eigenvalues of the Hessian ${\bf H}$ are $(\{ \omega_i^2\}, i=1,3N)$
representing  the squares of normal mode frequencies, and ${\bf W}({\bf r})$ 
are the corresponding eigenvectors. In a stable solid,  ${\bf r_0}$ can be
conveniently taken as the global  minimum of the potential energy
surface $U(R)$, which implies that ${\bf F}=0$ and ${\bf H}$ has
only positive eigenvalues corresponding to oscillatory modes.  The INM approach
for liquids  interprets ${\bf r}$ as the configuration at time $t$
relative to the configuration ${\bf r_0}$ at time $t_0$. Since typical
configurations, ${\bf r_0}$ are extremely  unlikely to be local minima,
therefore ${\bf F}\neq 0$ and
${\bf H}$ will have negative eigenvalues. The negative eigenvalue modes are
those which sample negative curvature regions of the PES,
including barrier crossing modes.
The ensemble-averaged INM spectrum, $\langle f(\omega) \rangle$,  is 
defined as
\begin{eqnarray}
f(\omega)= \left\langle \frac{1}{3N} \sum_{i=1}^{3N}
\delta(\omega-\omega_i)\right\rangle.
\label{eq:rho}
\end{eqnarray}
Quantities that are convenient for characterizing the instantaneous normal
mode spectrum are: (i) the fraction of imaginary frequencies ($F_{i}$), defined as
\begin{eqnarray}
F_{i}=\int_{im}^{} f({\omega}) d\omega
\label{eq:Fi}
\end{eqnarray}
and the Einstein frequency ($\omega_E$),  given by
\begin{eqnarray}
\omega_E^2&=&\int_{}^{}\omega^2  f(\omega) d\omega \nonumber \\
&=& \frac{\langle Tr {\bf H} \rangle}{m(3N-3)} 
\label{eq:omegaE}
\end{eqnarray}
where the  integral is performed over the entire range of frequencies, real as well as imaginary.

\section{Results}
\label{results}

\subsection{Phase Diagram}

\begin{figure}[ht]
  \begin{centering}
    \begin{tabular}{cc}
      \includegraphics[clip=true,width=6cm]{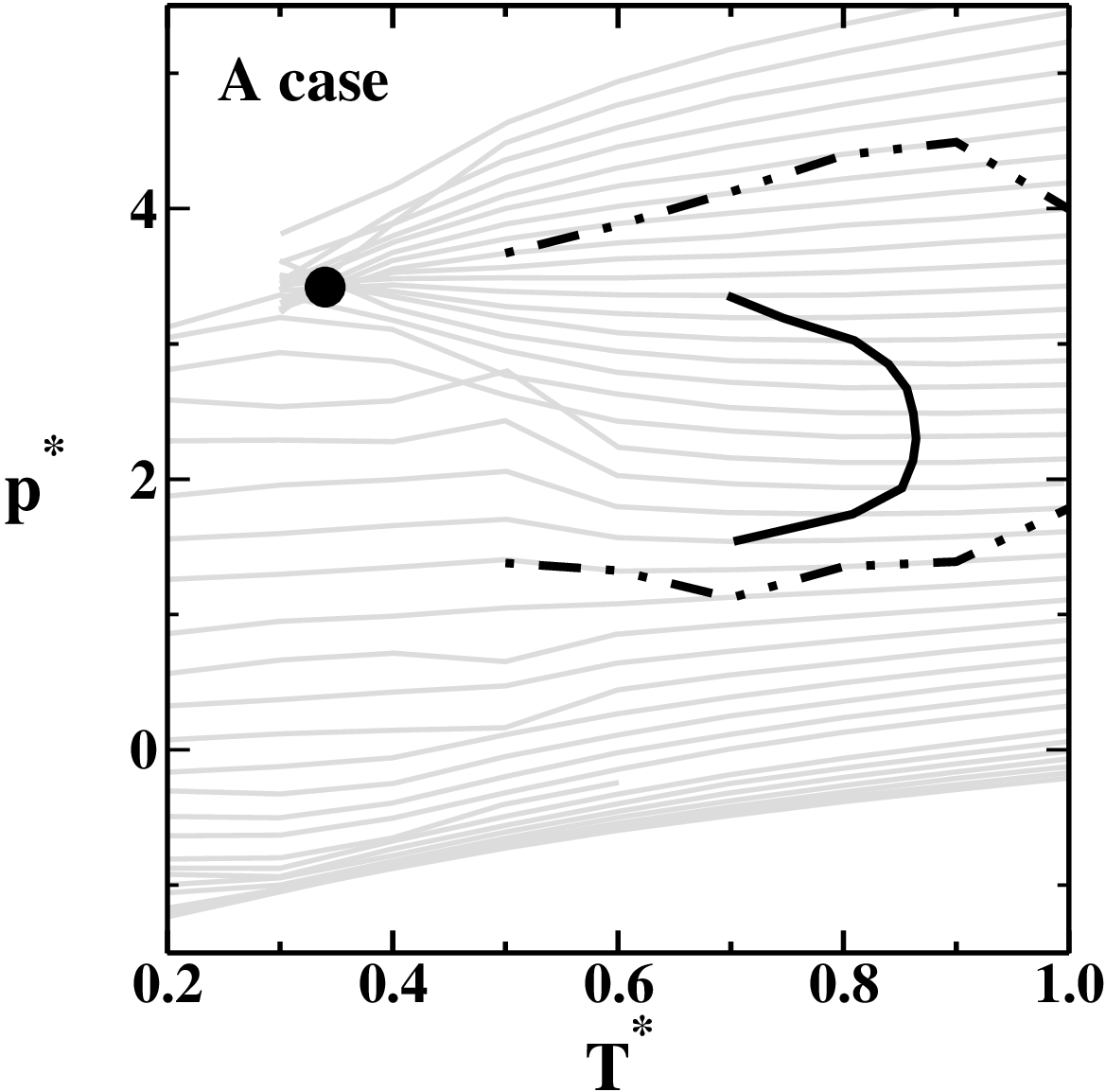} & \includegraphics[clip=true,width=6cm]{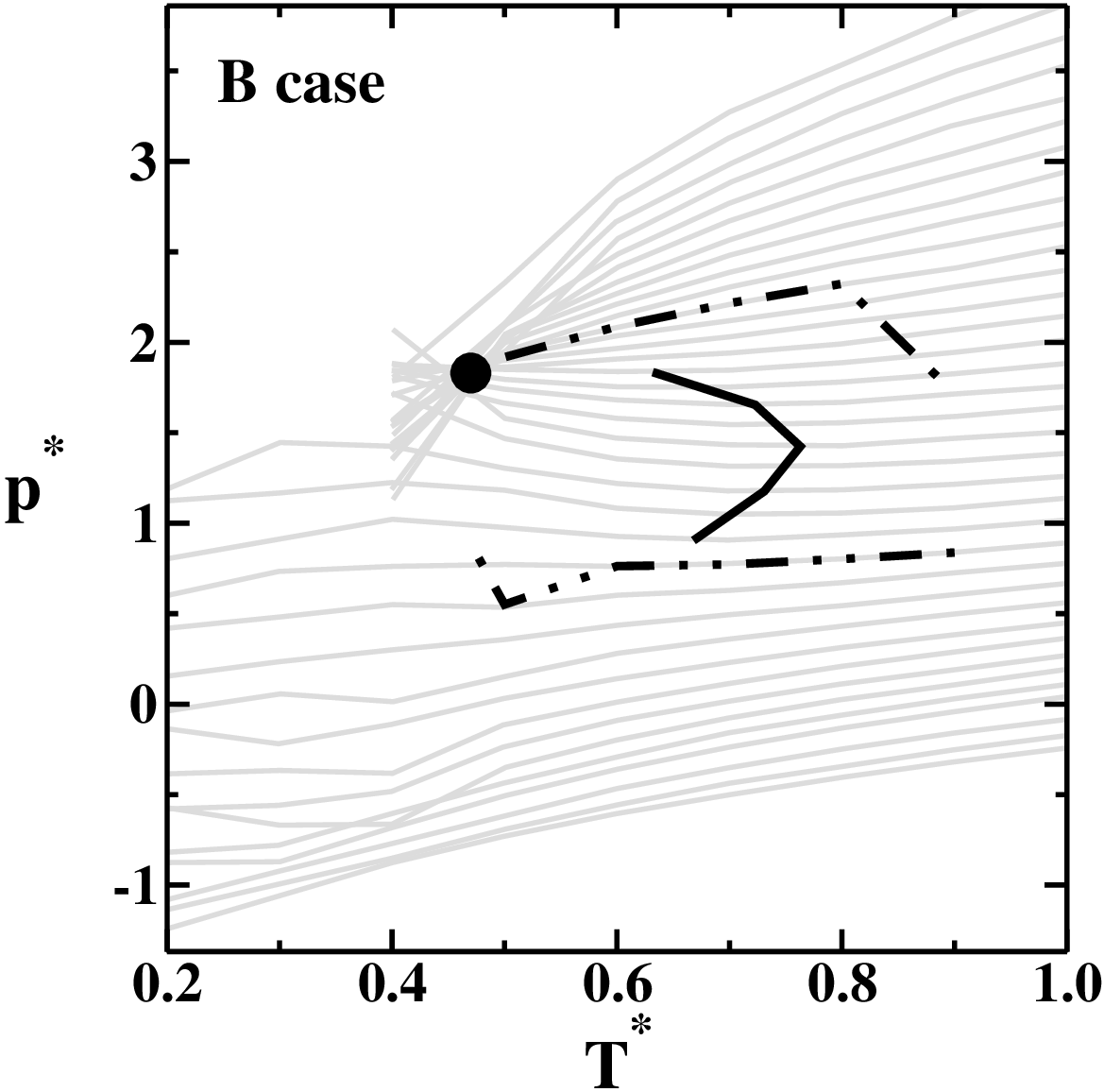}  \tabularnewline
      \includegraphics[clip=true,width=6cm]{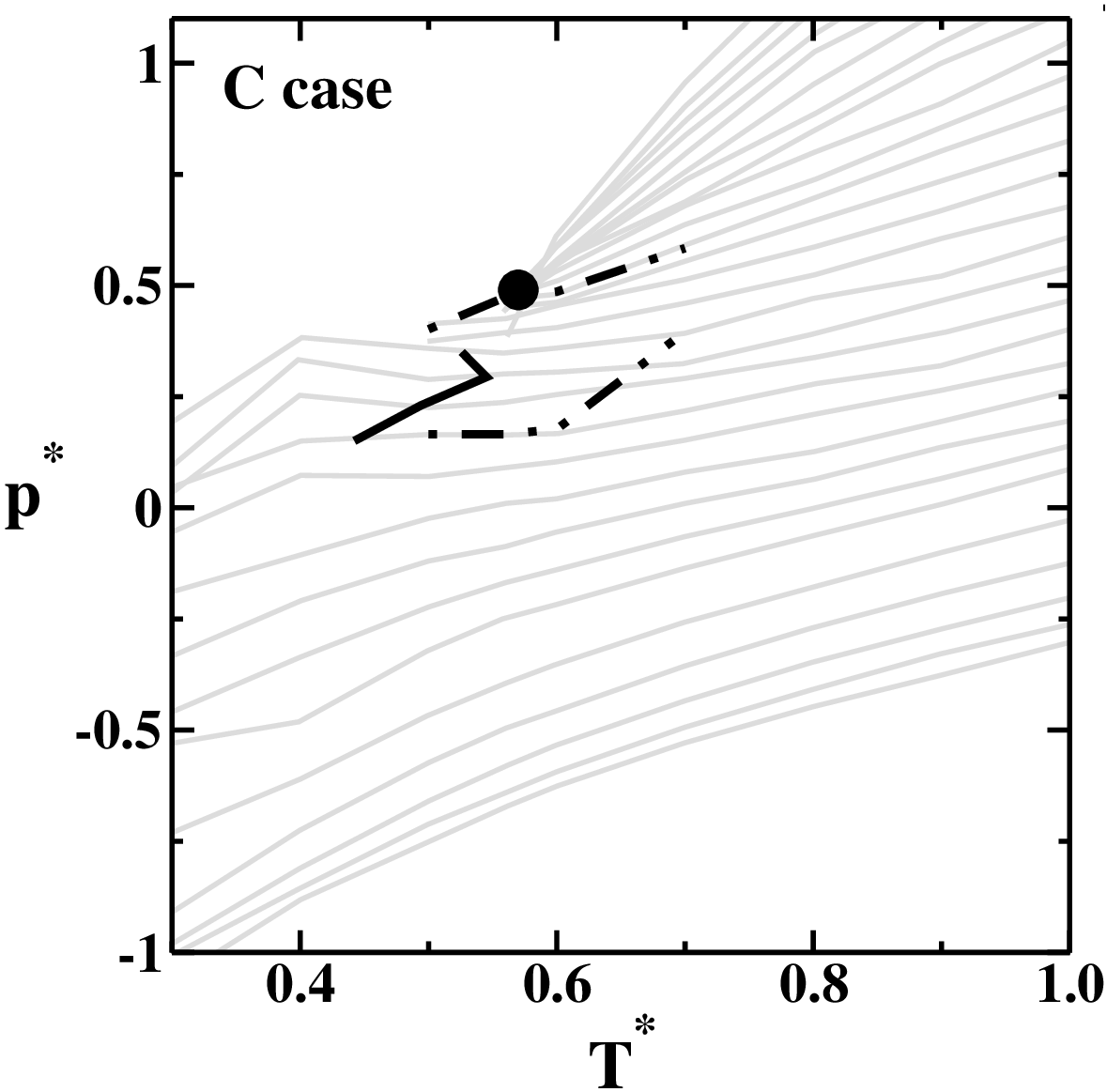} & \includegraphics[clip=true,width=6cm]{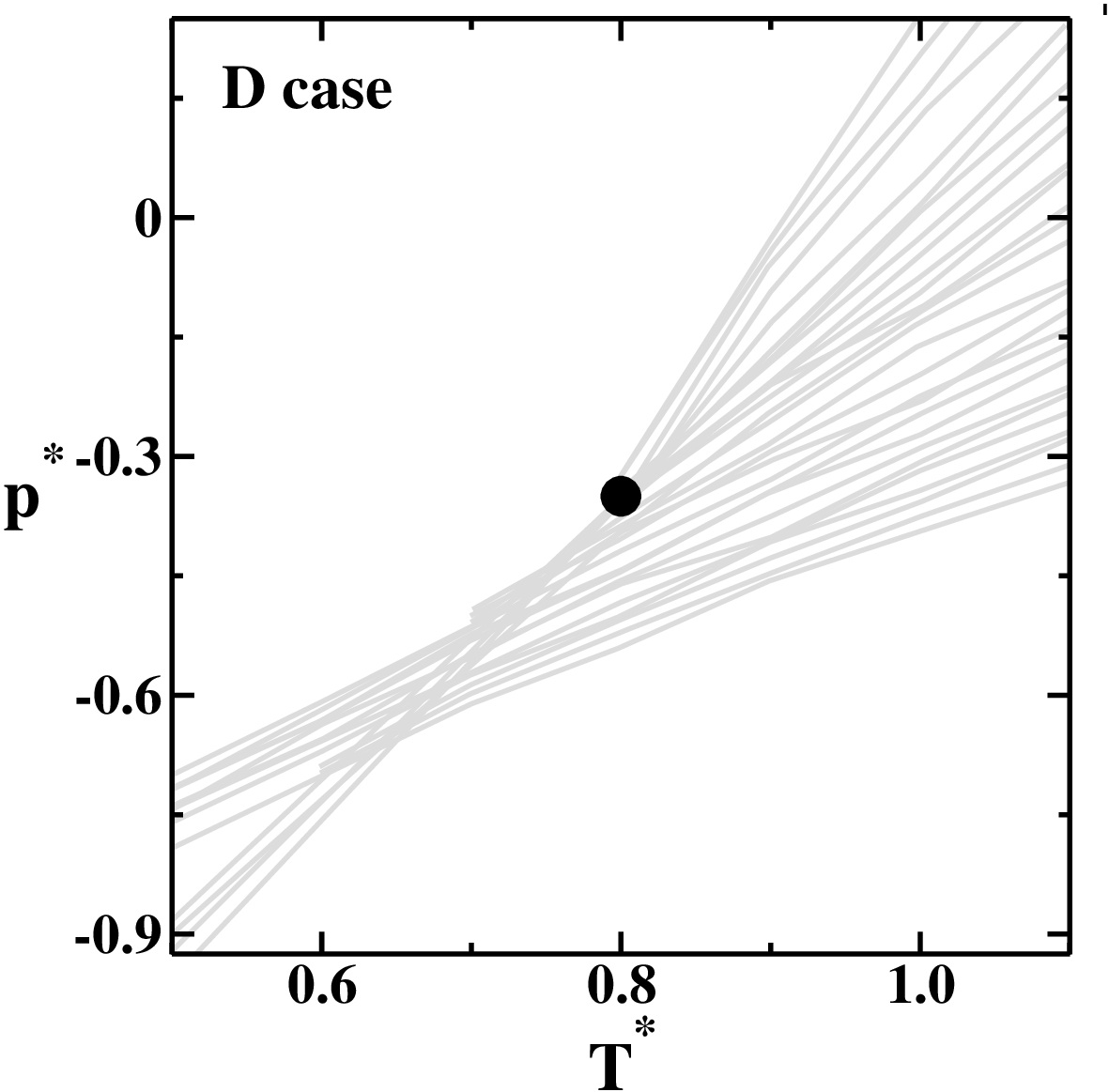} \tabularnewline
    \end{tabular}
    \par
  \end{centering}
\caption{Pressure-temperature phase diagram  for cases $A$, $B$, $C$ and 
$D$. The 
thin solid
lines are the isochores $0.30<\rho^*<0.65$. The liquid-liquid critical 
point is 
the  dot, the locus of temperatures of maximum density is the solid thick line
and  the locus of
diffusion extrema is the dot-dashed line.}
\label{fig:PT} 
\end{figure}

Figure~\ref{fig:PT} illustrates the pressure-temperature phase diagram for the four cases of the potential~\cite{Ba09}.
Due to the presence of the attractive interaction, all four cases have a liquid-gas transition with an associated critical point that
is not shown here. In addition, all the four model liquids studied here  have a liquid-liquid critical point. Cases A, B and C have water-like 
density and diffusional anomalies. The solid bold lines represent the locus of temperatures of  maximum density (TMD) for different isobars.
State points enclosed by the TMD locus represent the regime of density anomaly within which $(\partial\rho /\partial T)_P > 0$.
The maximum temperature along the TMD locus, denoted by $T^{max}_{TMD}$, is the threshold temperature for onset of the density anomaly.
 The dot-dashed lines are the temperatures of maximum and minimum diffusivity along different isotherms~\cite{Ba09}.

\begin{figure}[h]
  \begin{centering}
    \begin{tabular}{cc}
      
      \includegraphics[clip=true,width=7cm]{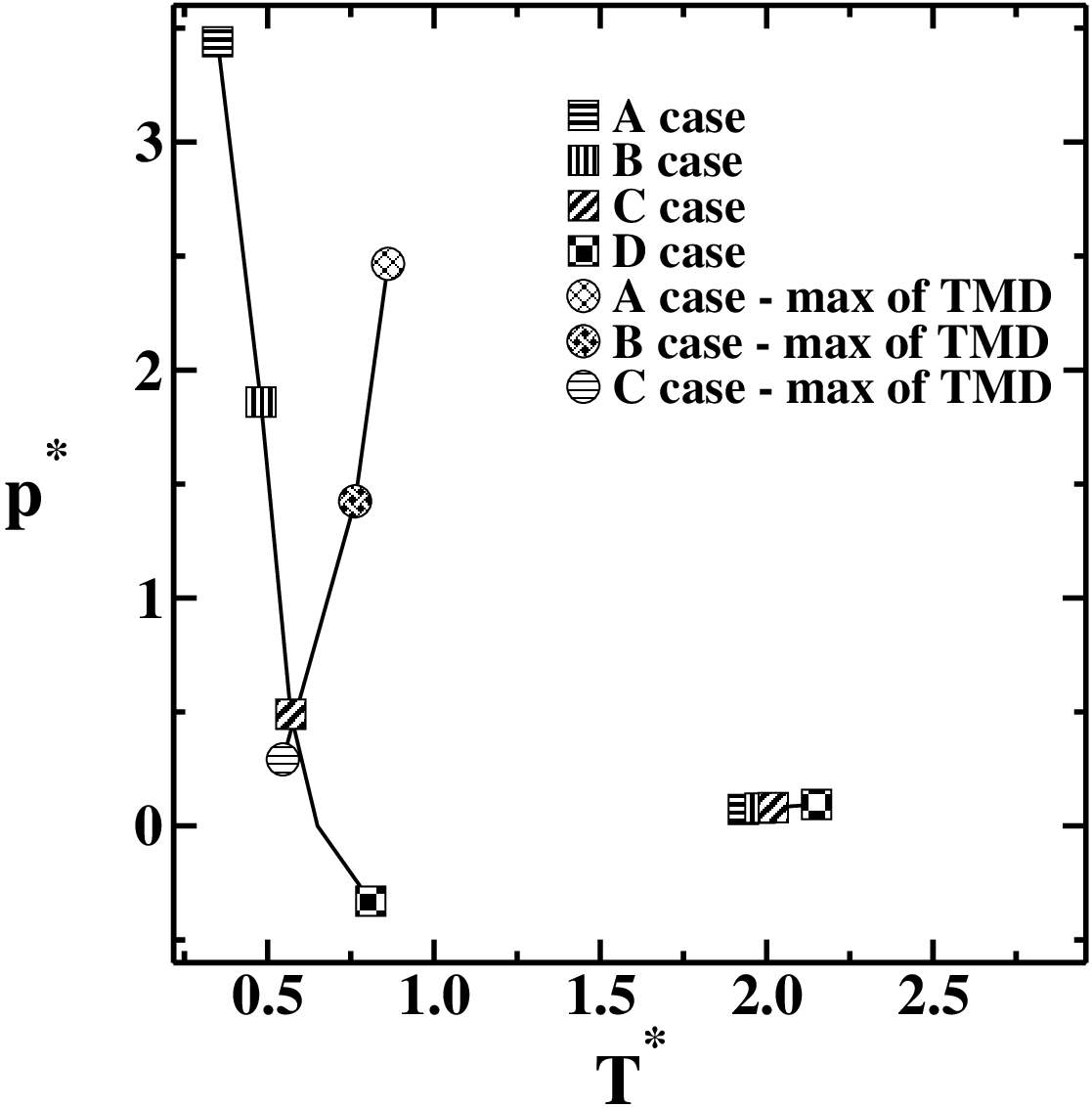}
    \end{tabular}
    \par
  \end{centering}
\caption{Pressure versus temperature locations of the critical points for the potentials $A-D$.}
\label{fig:PT-CP} 
\end{figure}

This overall change in the nature of the liquid-state phase diagram for the four multi-Gaussian liquids is summarized in Figure~\ref{fig:PT-CP}.
Clearly, as the second length scale shifts from an inflexion point on the repulsive shoulder to a well with progressively increasing depth and curvature, the region
of liquid state anomalies shrinks and disappears.  The figure   illustrates how  the pressure and temperature associated 
the liquid-gas and liquid-liquid critical points vary with the potentials $A,B,C$ and $D$. The same graph also shows that as the shoulder becomes  deeper,  the maximum temperature of the
TMD locus, which marks the onset temperature for thermodynamically anomalous behaviour, approaches the liquid-liquid critical point. 

Since the thermodynamic and mobility anomalies of water are
correlated, we first focus on understanding the thermodynamic
condition for the presence of density anomaly. This may be stated
as:
\begin{equation}
\frac{\partial S}{\partial \rho}= -\frac{V^2\alpha}{N K_T}>0 
\label{eq:ds/drho}
\end{equation}
where $\alpha$ is the thermal expansion coefficient and  $K_T$ is the isothermal compressibility.
For the system to have a large anomalous region, the ratio $\alpha/K_T$ should be therefore large
and negative. Near the critical point, the compressibility, $K_T$, and thermal expansion coefficient, $\alpha_T$,   diverge, however
the compressibility diverges with a large exponent making the ratio zero. In this case, the 
 condition given by Eq.~(\ref{eq:ds/drho}) can  not be fulfilled. This suggests that 
near the liquid-liquid critical point  the system prefers to undergo a phase separation into  high- and low-density liquids, rather than  show a smooth entropy anomaly.

\subsection{Excess Entropy and Pair Correlations}

As discussed in the previous section, the density anomaly 
corresponds to a set of state points for which 
$(\partial S /\partial\rho )_T > 0$.  The total
entropy is a sum of the ideal ($S_{id}$)  and excess ($S_{ex}$)
contributions. Since $S_{id}$ decreases monotonically with
increasing density, therefore a density anomaly must imply
the presence of an excess entropy 
anomaly,  $(\partial S_{ex} /\partial\rho )_T > 0$.
Errington \emph{et al.} have further  shown that the 
strength of the excess entropy anomaly required to give rise to  
density anomaly is given by the condition  
$\Sigma_{\mathrm{ex}}= \left(\partial (S_{\mathrm{ex}}/Nk_B) /\partial 
\ln \rho\right)_T>1$~\cite{Er06}. By approximating the excess entropy with the  
two-body correlation contribution of $s_2$ [see Eq.~(1)],
we  relate the structural information in the radial distribution 
function of the fluid to the thermodynamic behaviour.

\begin{figure}[htbp]   
\centering   
\includegraphics[clip=true,scale=0.32]{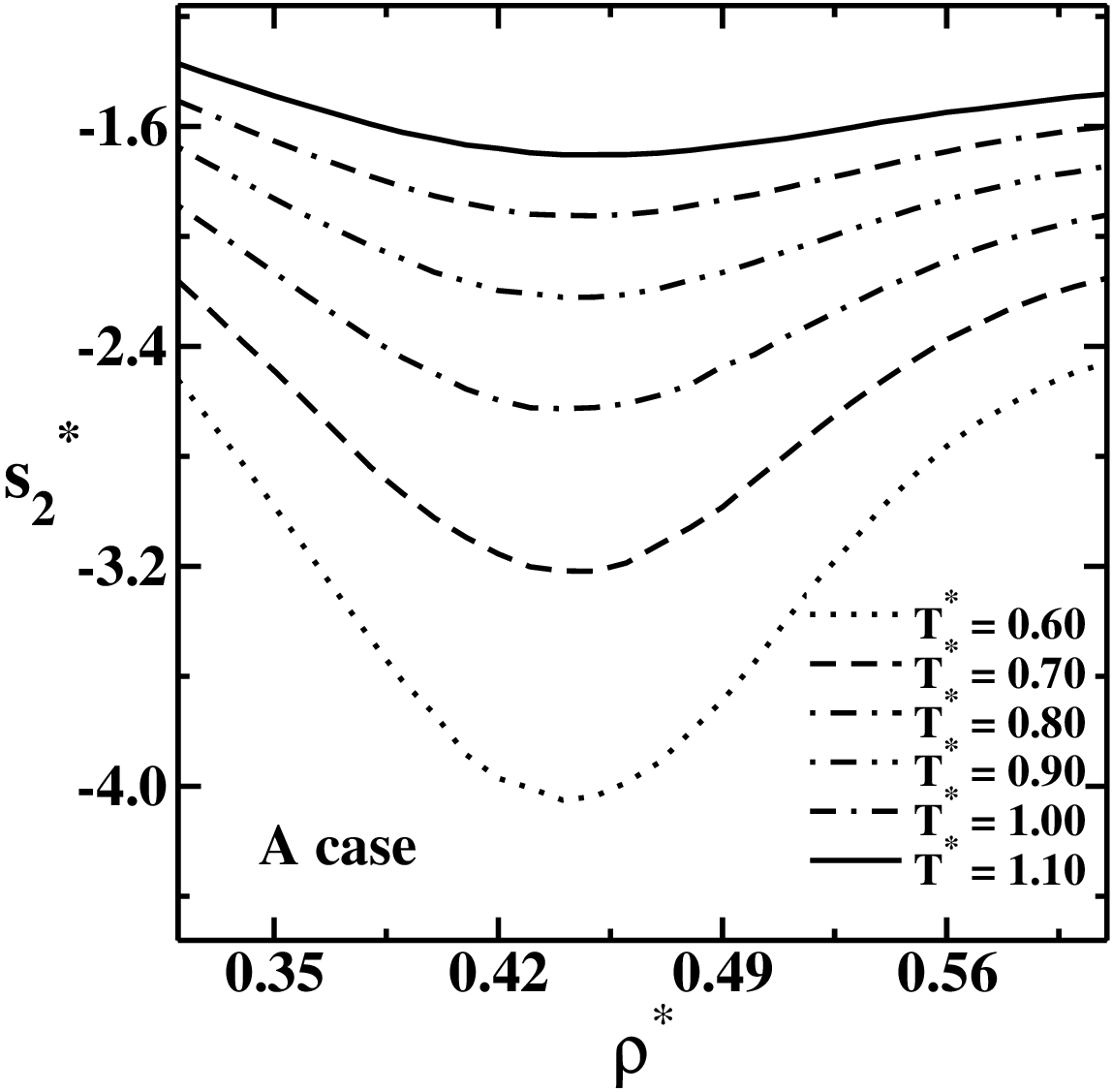}
\includegraphics[clip=true,scale=0.32]{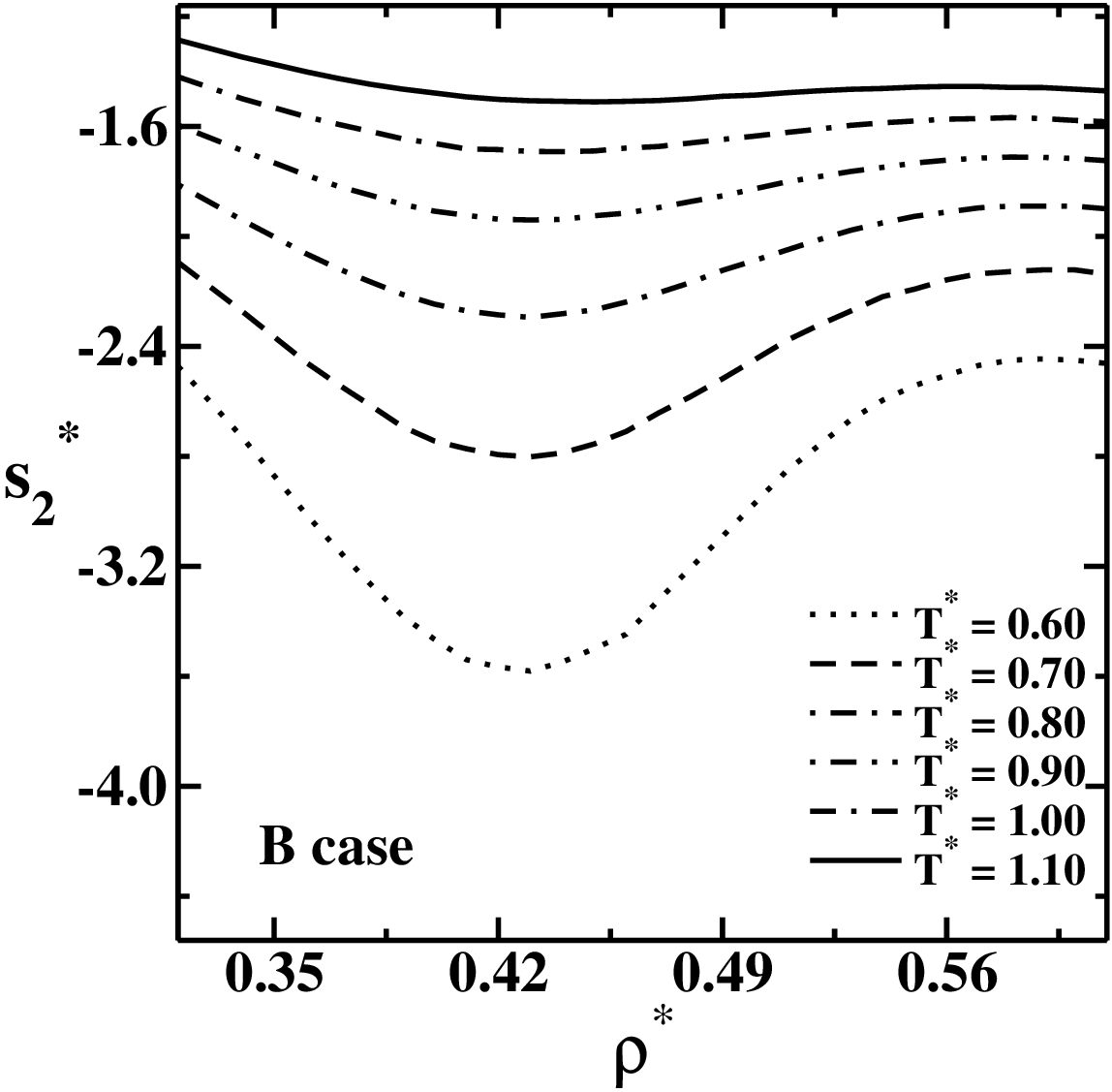}
\includegraphics[clip=true,scale=0.32]{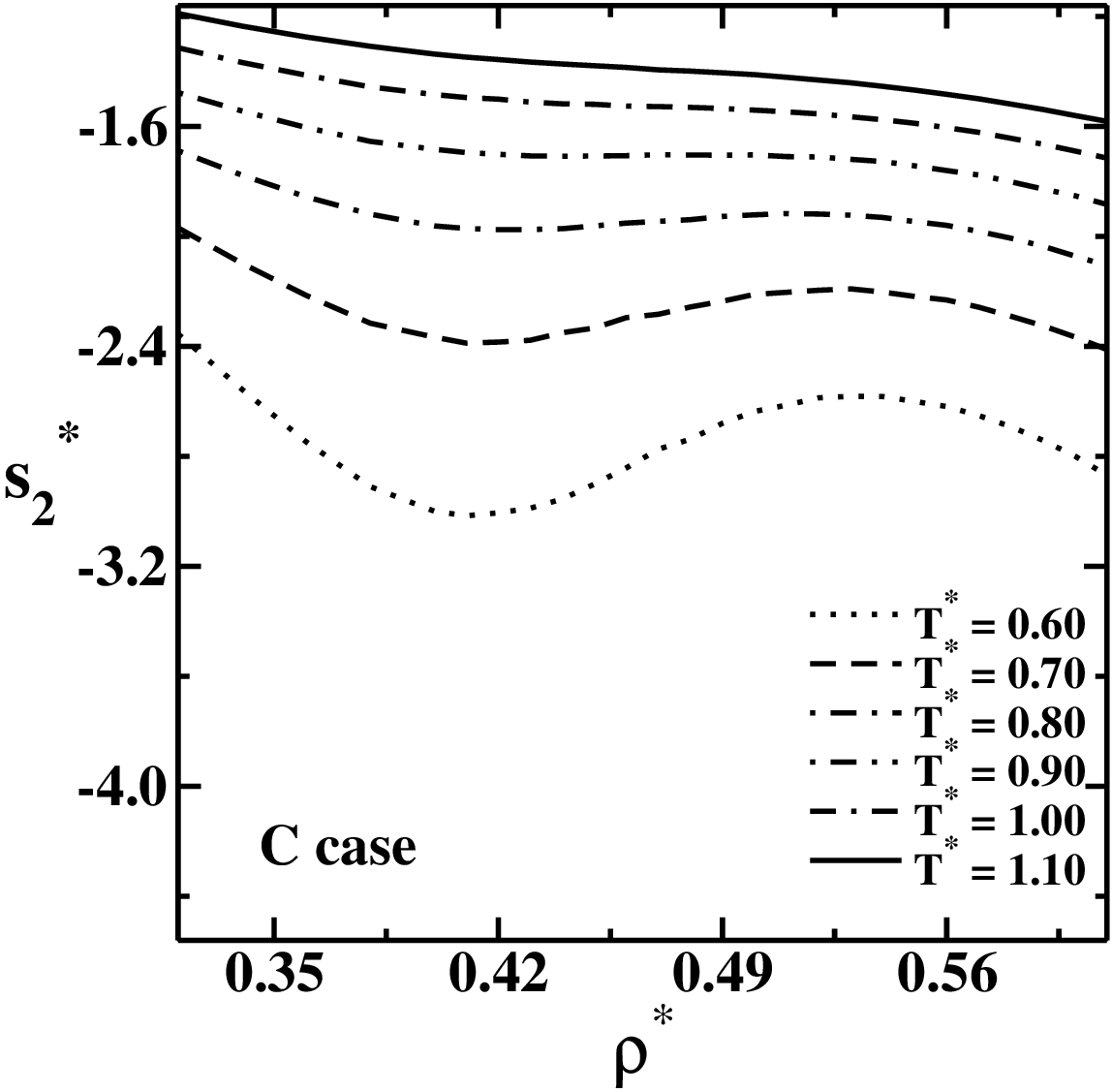}
\includegraphics[clip=true,scale=0.32]{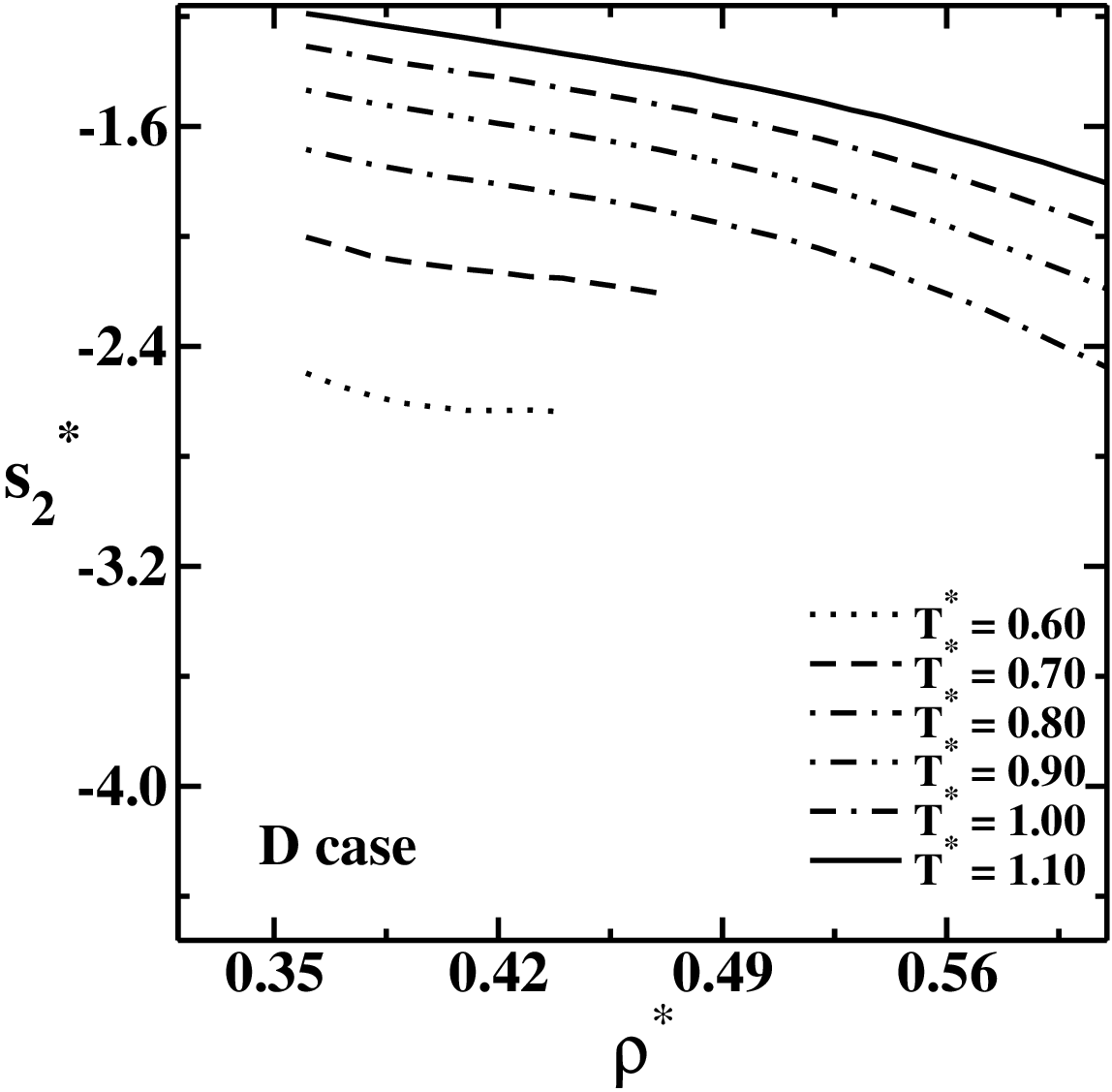}
\caption{Pair entropy versus density for  for the cases
$A$, $B$, $C$ and $D$ for various temperatures.}
\label{fig:s2-sex}
\end{figure}

\begin{figure}[htbp]   
\centering   
\includegraphics[clip=true,scale=0.5]{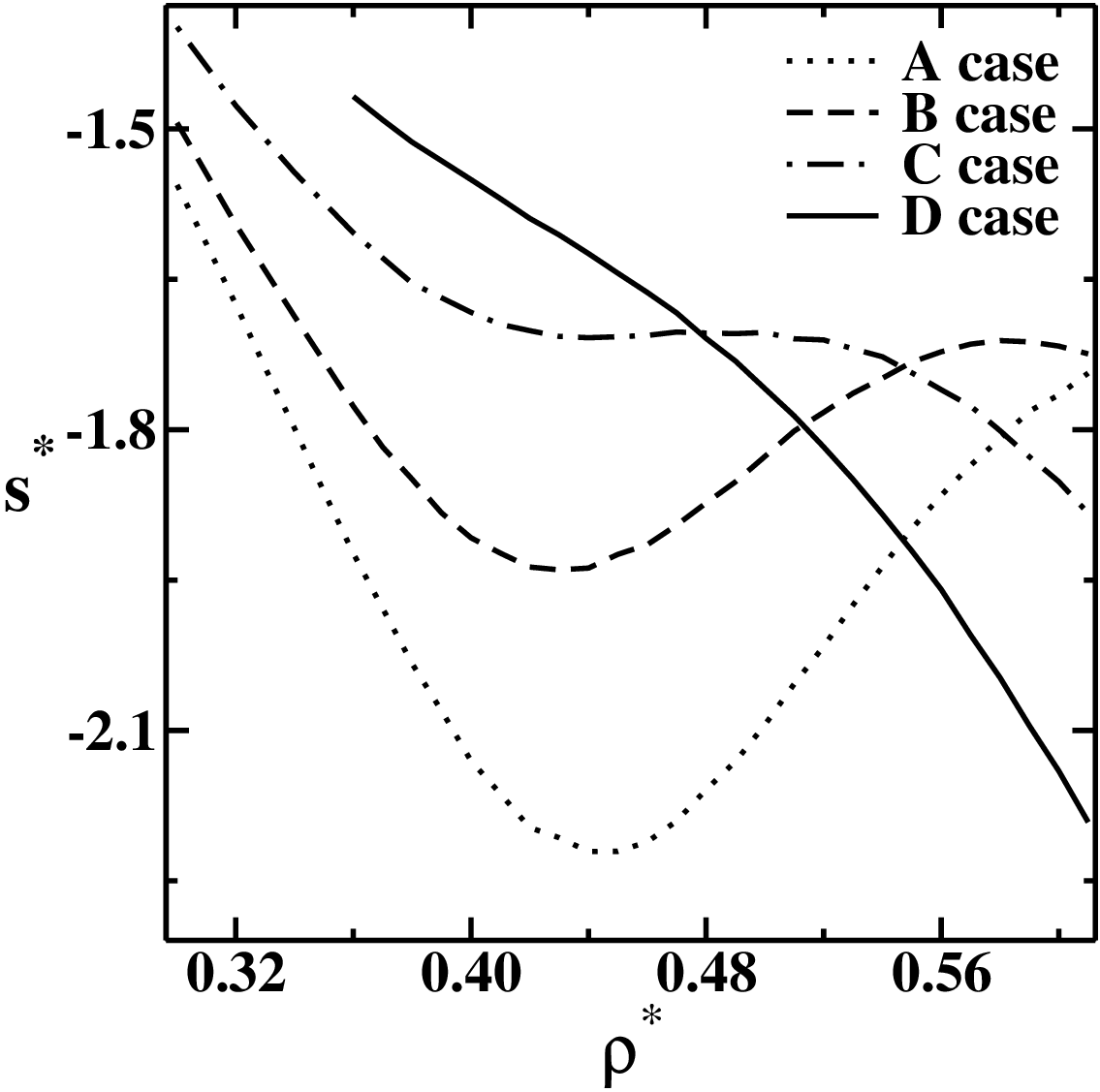}
\caption{ Pair entropy versus density for the cases
$A$, $B$, $C$ and $D$ at  $T^*=0.9$. }
\label{fig:s2-sex-T-all}
\end{figure}

Figure~\ref{fig:s2-sex} illustrates the  $s_{2}^*(\rho)=S_2/Nk_B$ versus $\rho^*$ for
various temperatures and for the potentials $A,B,C$ and $D$. For  cases 
$A,B$ and $C$, at low temperatures, there is a rise in excess entropy 
on isothermal compression   characteristic of water-like liquids~\cite{scc06,ac09pre,agc09,Mi06b,oscb10} that
contrasts with the behaviour of simple liquids where free volume arguments
are sufficient to justify a monotonic decrease in entropy on
isothermal compression. For  case $D$,  no anomaly is observed in the 
pair entropy
even at low temperatures. The progressive attenuation of the anomalies
on going from case $A$ to case $D$, is illustrated in 
Figure~\ref{fig:s2-sex-T-all}
which compares the behavior of the pair entropy versus density
for all studied potentials at a given temperature, $T^*=0.9$.
 This graph together with Figure~\ref{fig:PT-CP}
indicates that as the maximum temperature at the
TMD line approaches the liquid-liquid critical 
temperature, the pair entropy curve becomes 
more flat and the anomalous behavior disappears.

The origin of the pair entropy anomaly  in fluids with
two length scales can be explained in terms of a 
competition between two length scales at intermediate densities. Only a single length scale dominates in the low- and high-density limits while at intermediate densities, where
both length scales are present, can be regarded as quasi-binary systems with a mixing entropy. The radial distribution functions shown in our previous study clearly demonstrate
the presence of two length scales. They also show that with increasing  temperature, the shorter length scale peak of $g(r)$ becomes more prominent in cases $A, B$ and $C$. In
contrast, in case $D$, both length scales associated with the first and second peak of the $g(r)$ broaden with increasing temperature as a consequence of which there
is no emergence of an anomaly with decreasing temperature.

The crucial question to ask in the multi-Gaussian family of
water models is why, despite the presence of two length
scales at intermediate densities, the pair entropy anomaly
is progressively lost as the shoulder goes from being an inflexion point to a minimum with about twice the depth 
as the outer, attractive well. Clearly the rise in entropy
with isothermal compression due to mixing of two length scales
is counteracted by additional effects. To understand this
we note that the entropy of an one-dimensional harmonic oscillator of frequency $\omega$ is given by: 
\begin{equation}
\frac{s_\omega}{Nk_B} = 1 -\ln (\beta\hbar\omega ) \; .
\end{equation}

The increasing curvature of the short-range minimum,
relative to  the attractive minimum, implies that a pair separated 
trapped in the
shoulder minimum will have lower vibrational entropy than one trapped in the 
broad shallow attractive minimum. As a consequence, at intermediate 
densities, while
the presence of two length scales will increase entropy, the loss of entropy
when the pairs are located in the short-range minimum will tend to 
decrease entropy.
As the shoulder minimum becomes deeper, the second effect becomes 
more important
and the excess entropy anomaly disappears. In systems such as the two-scale
linear ramp, such curvature-dependent effects will be absent.

It is also interesting to consider the shifting of the liquid-liquid critical
point to lower pressures and higher temperatures. For a temperature-driven
phase separation into low-density liquid (LDL) and high-density liquid (HDL), increasing
energetic stabilization of one length scale relative to the other is
required. In case A, this is clearly due to the outer attractive well
with a depth of about $\Delta U^* \approx 0.3$  and
$T_c^* \approx 0.3$.  In case D, this is due to the shorter
length scale with a well depth of about $\Delta^*=1.00$ and 
$T_c^* \approx 0.8$. For a shallow shoulder, the high
density liquid is stabilized under high pressure. Within the HDL phase
particles are occupying the shoulder scale.  The density 
anomalous region, characterized by having particles in the two scales 
occurs at the pressure range of the 
low density liquid phase.  As the shoulder scale
becomes deeper, less pressure is needed to form the high density
liquid. The LDL phase occupies a smaller pressure range
and therefore the density anomalous region shrinks. For a 
very deep shoulder well as in case D, the HDL requires no
pressure to be formed, the LDL is at negative pressures
and the anomalous density regime disappears.

\subsection{Diffusion and Rosenfeld Reduction Parameter}

\begin{figure}[htbp]   
\centering   

\includegraphics[clip=true,scale=0.3]{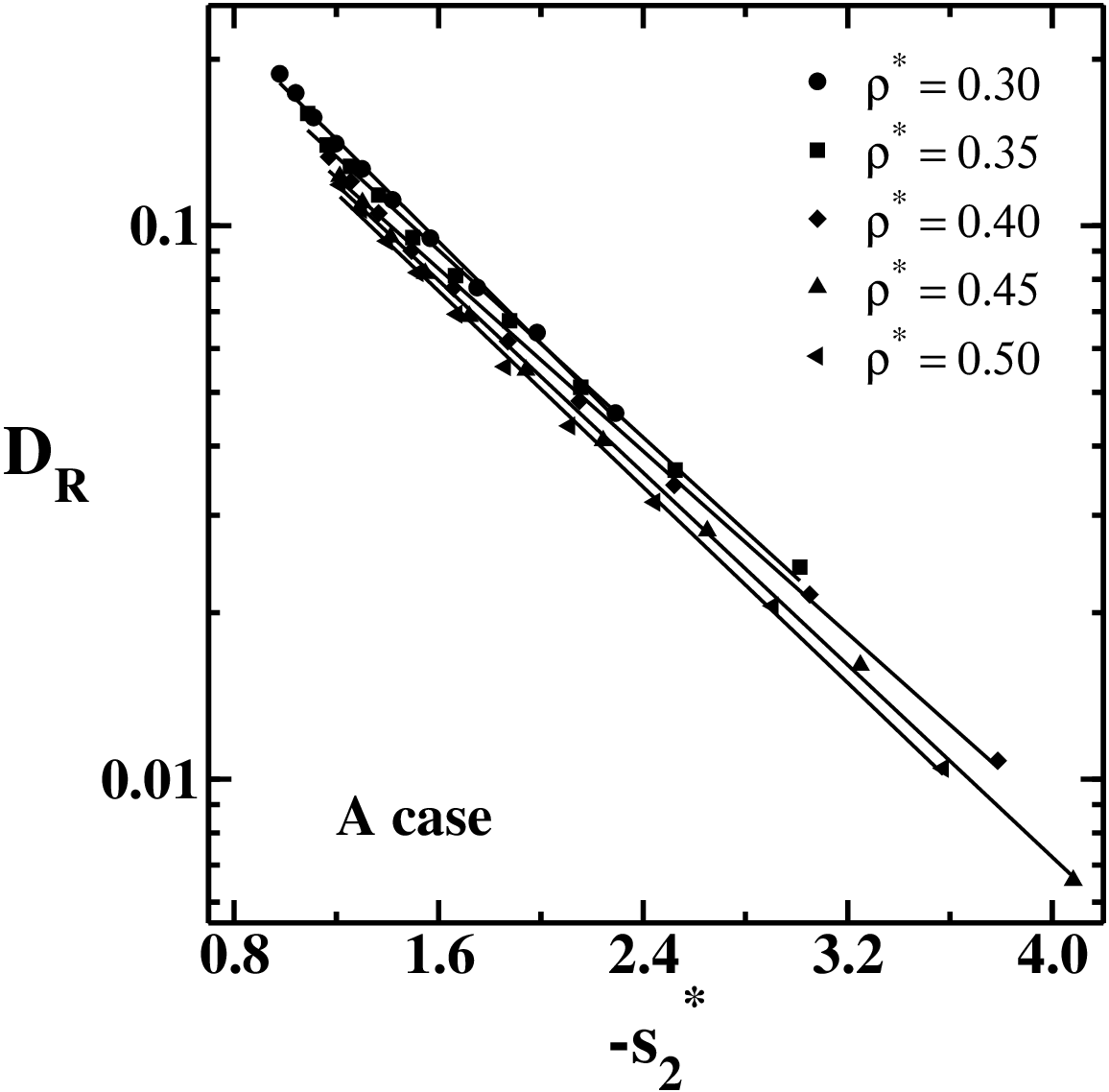}
\includegraphics[clip=true,scale=0.3]{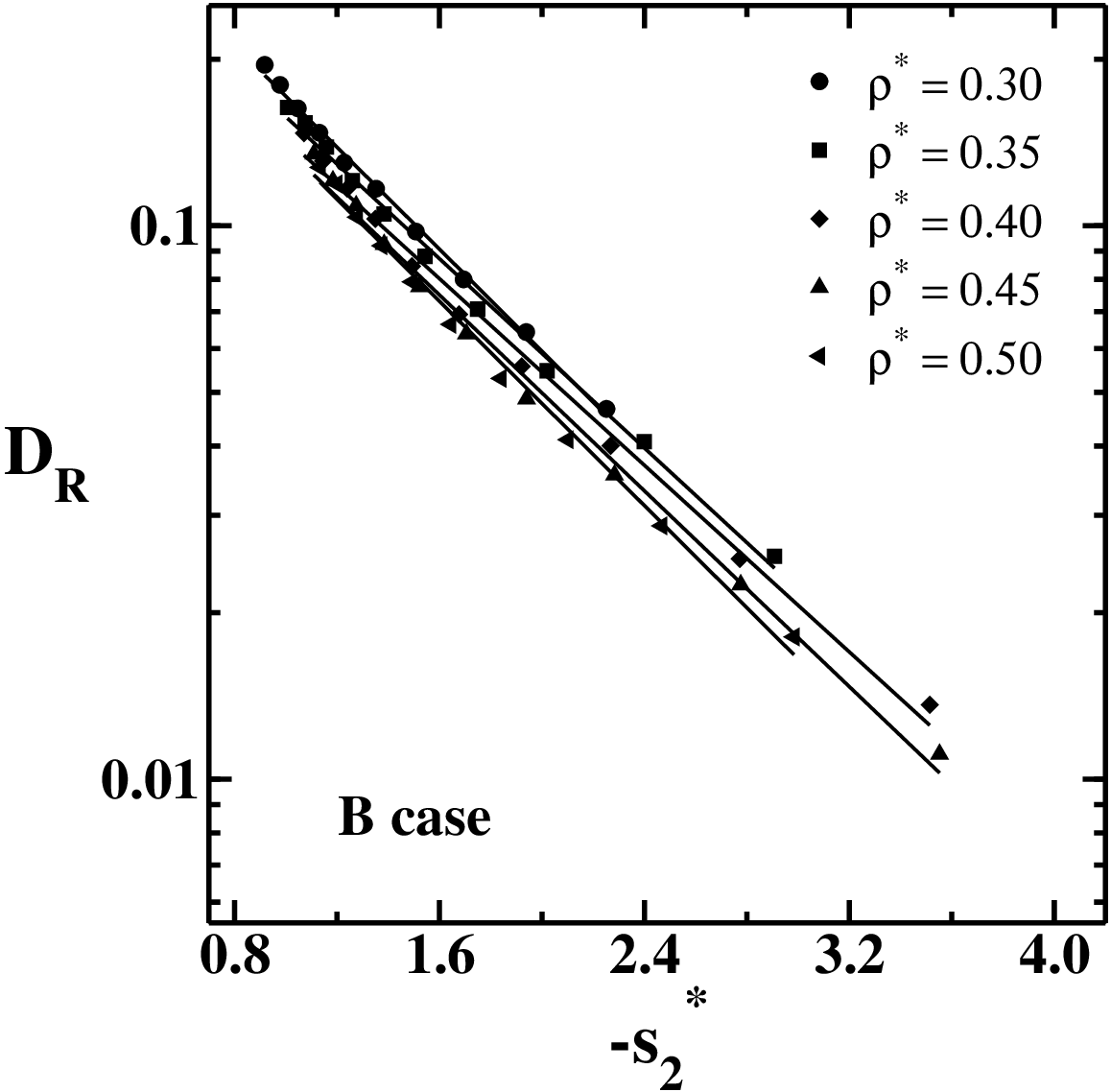}
\includegraphics[clip=true,scale=0.3]{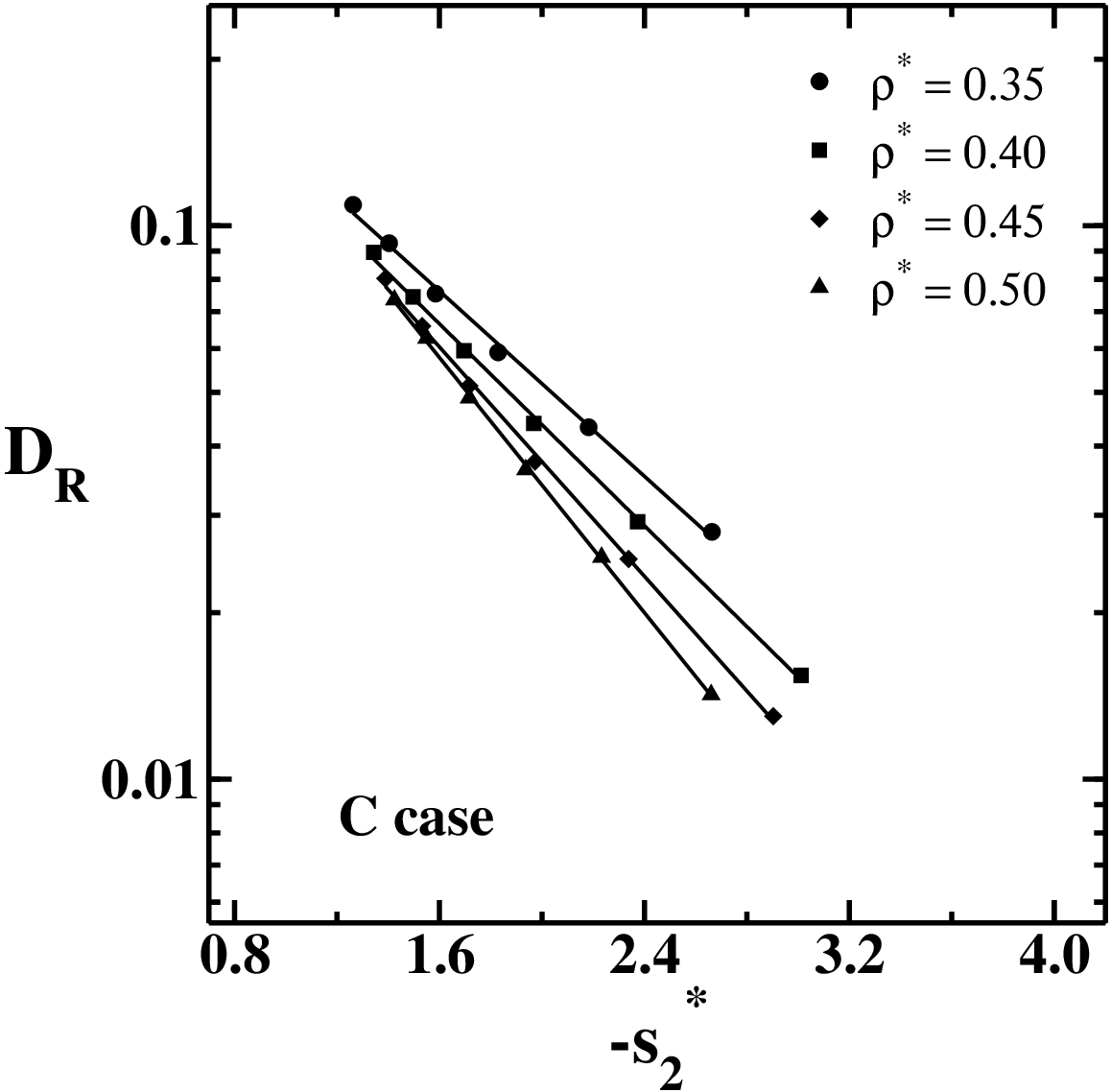}
\includegraphics[clip=true,scale=0.3]{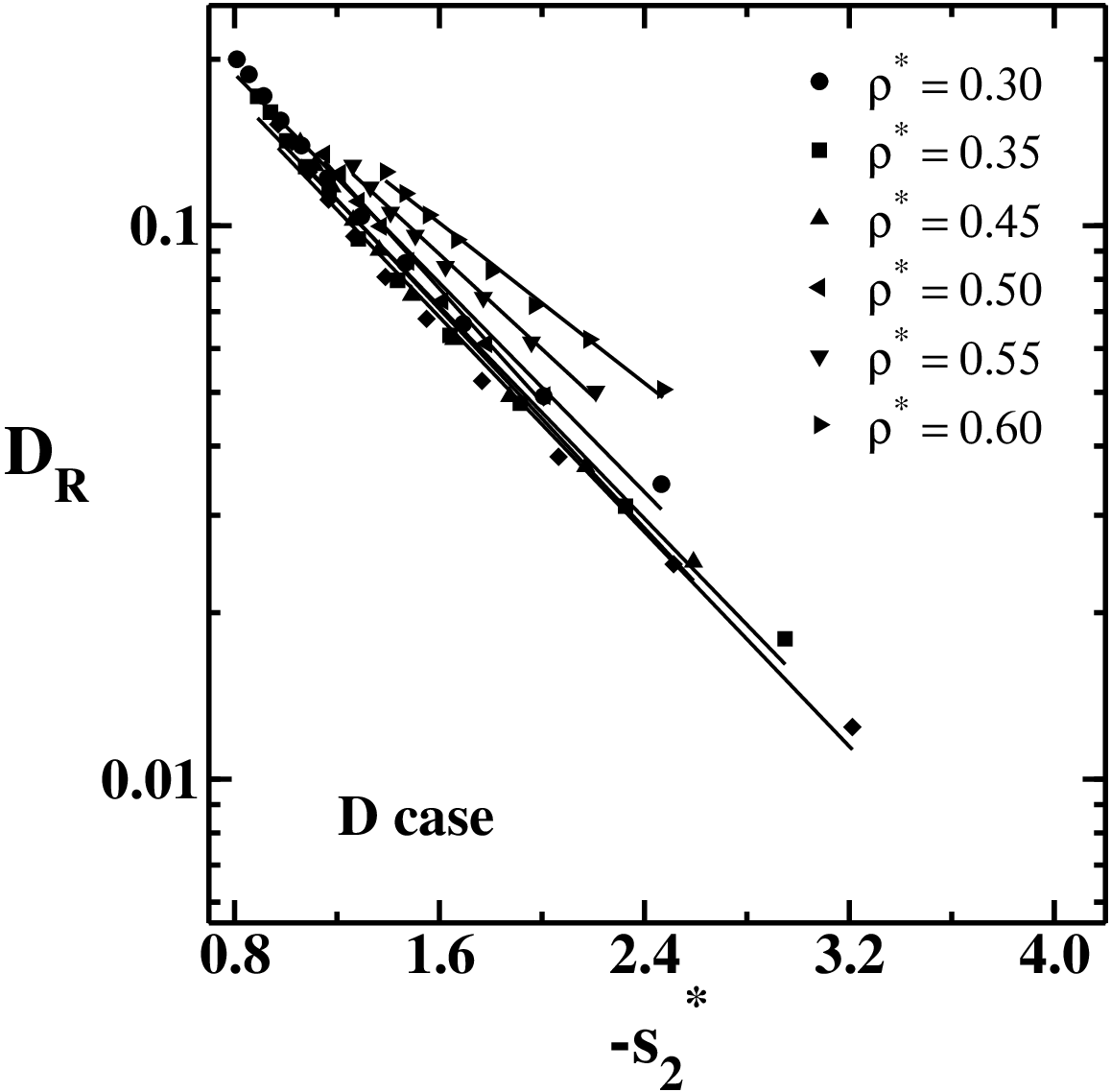}
\caption{Diffusion in Rosenfeld units as a function of $-s_{2}^*$ for the cases $A, B, C$ and $D$.  }
\label{fig:D-DR}
\end{figure}

Previously we have shown that the diffusion coefficient in the cases $A,B$ and $C$ decreases with 
the decrease of the density for a certain range of densities~\cite{Ba09}. The region
in the pressure temperature phase diagram limited by the maxima and minima of
the diffusion coefficient is  illustrated as dot-dashed  lines in Fig.~\ref{fig:PT}.   In the case
$D$ the diffusion coefficient increases with  the decrease of the density as in normal liquids.
It is interesting to notice that the same behavior is also observed in the pair entropy
suggesting that the anomalies present in these two quantities might be related. In order
to check this hypothesis we now consider the scaling relationship between the
diffusivity and the pair entropy.  Using the Rosenfeld macroscopic reduction 
parameters for the length  as $\rho^{-1/3}$ and the thermal velocity as $(k_BT/m)^{1/2}$, 
the dimensionless diffusivity is defined as
\begin{equation}
D_R \equiv  D\frac{\rho^{1/3}}{(k_BT/m)^{1/2}}\; .
\label{eq:D*sexc}
\end{equation}
The scaling of the reduced diffusivity, $D_R$ with pair entropy, $s_2^*$ is illustrated 
in Fig.~\ref{fig:D-DR}. Previous results for core-softened fluids~\cite{assac10} suggest
that  $\Delta S= S_{ex} -S_2$ tends to be density dependent in anomalous fluids, resulting in 
a stronger isochores dependence when $\ln(D_R)$ is plotted against $S_2$, rather than against
$S_{ex}$. In the present study, we have computed only $S_2$ and therefore Fig.~\ref{fig:D-DR}
shows scaling with respect to $S_2$. For case $A$, the collapse of data from all the state
points on a single line is quite good. As we progress from case $A$ to case $D$, the
isochores dependence of the scaling parameters becomes more pronounced suggesting that 
 the density dependence of $\Delta S$ increases on going from
case $A$ to case $D$. 

\subsection{The Instantaneous Normal Mode Spectrum}

\begin{figure}[htbp]
   \centering
   \includegraphics[clip=true,scale=0.35]{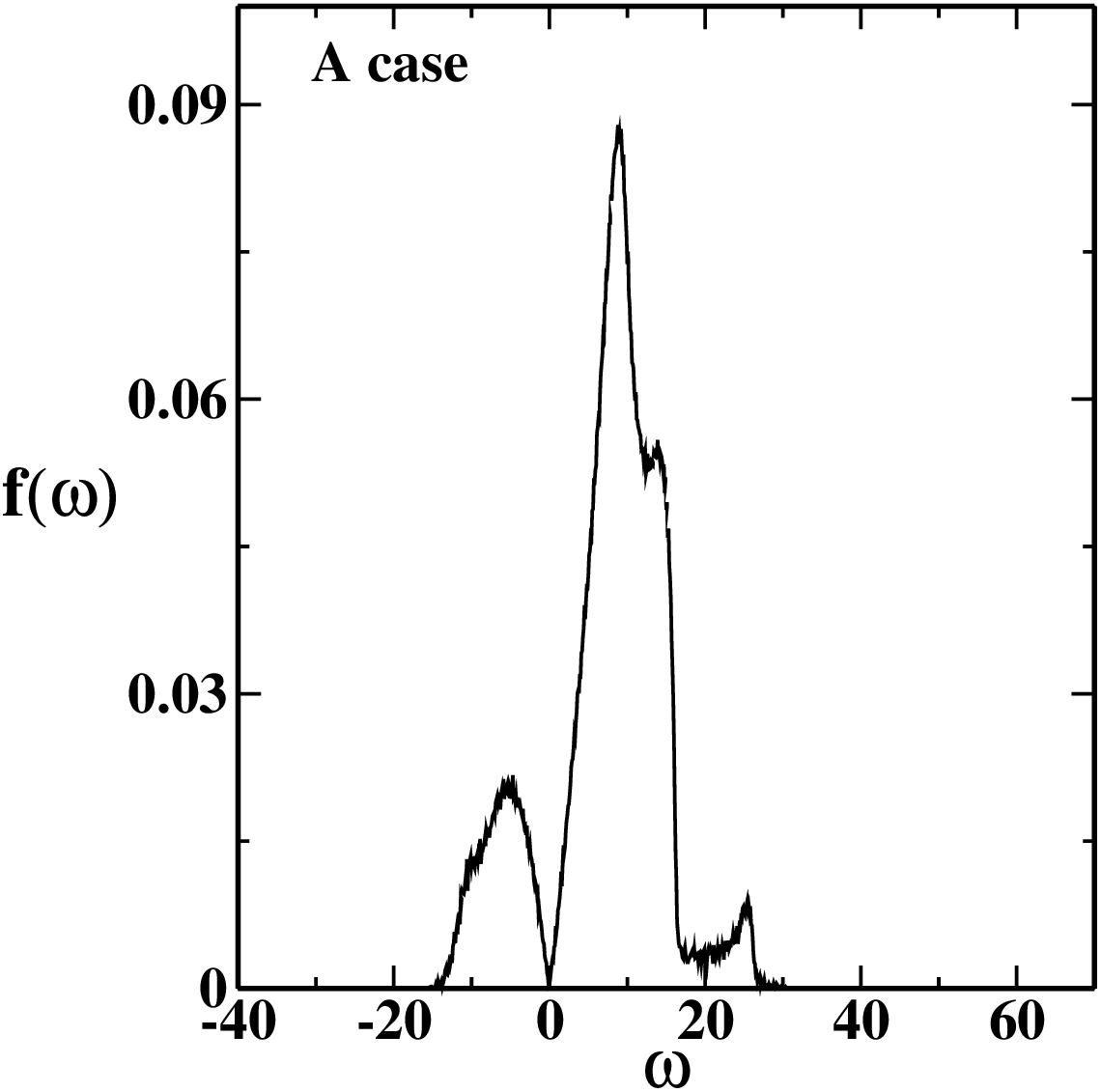}
   \includegraphics[clip=true,scale=0.35]{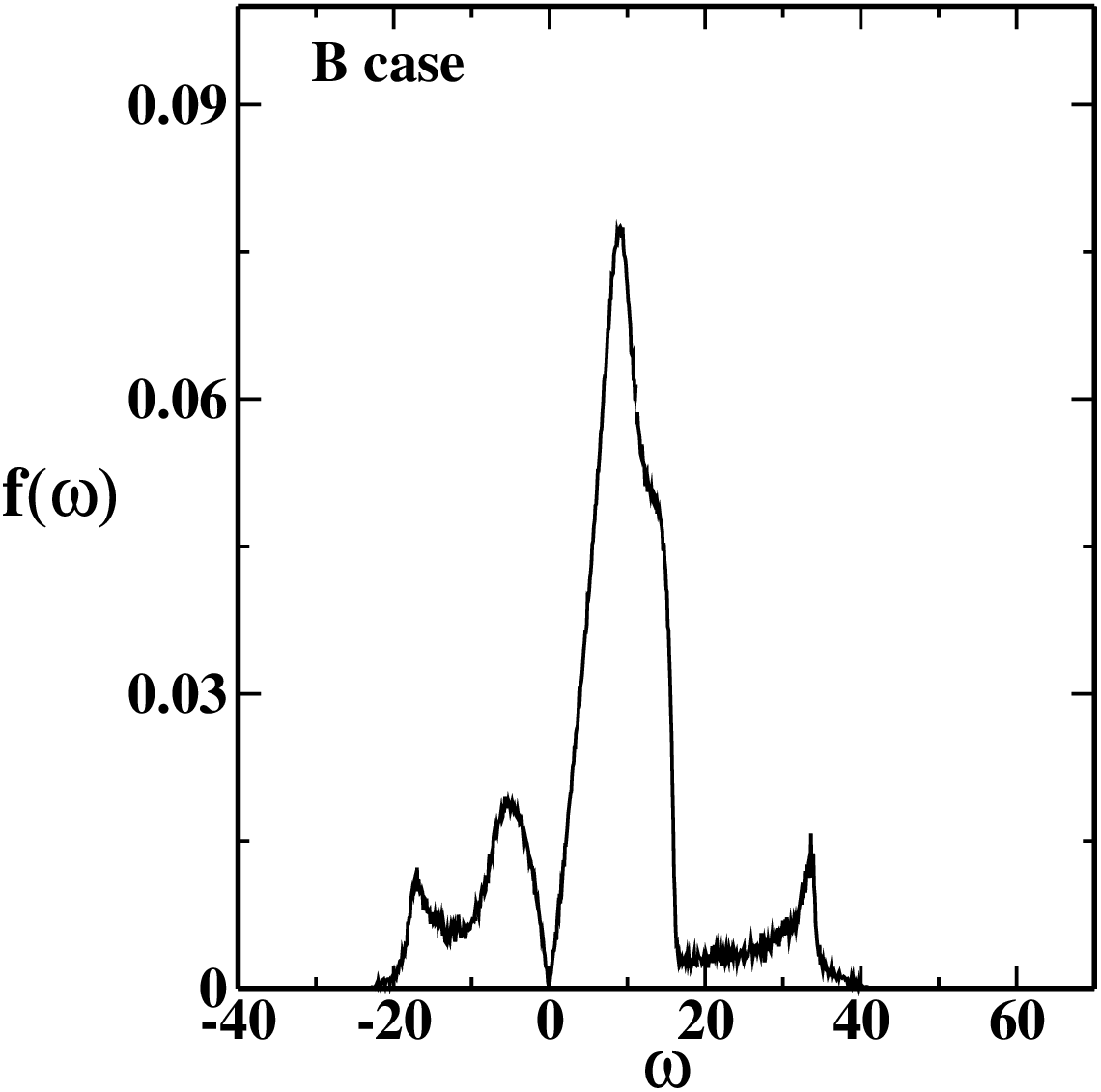}
   \includegraphics[clip=true,scale=0.35]{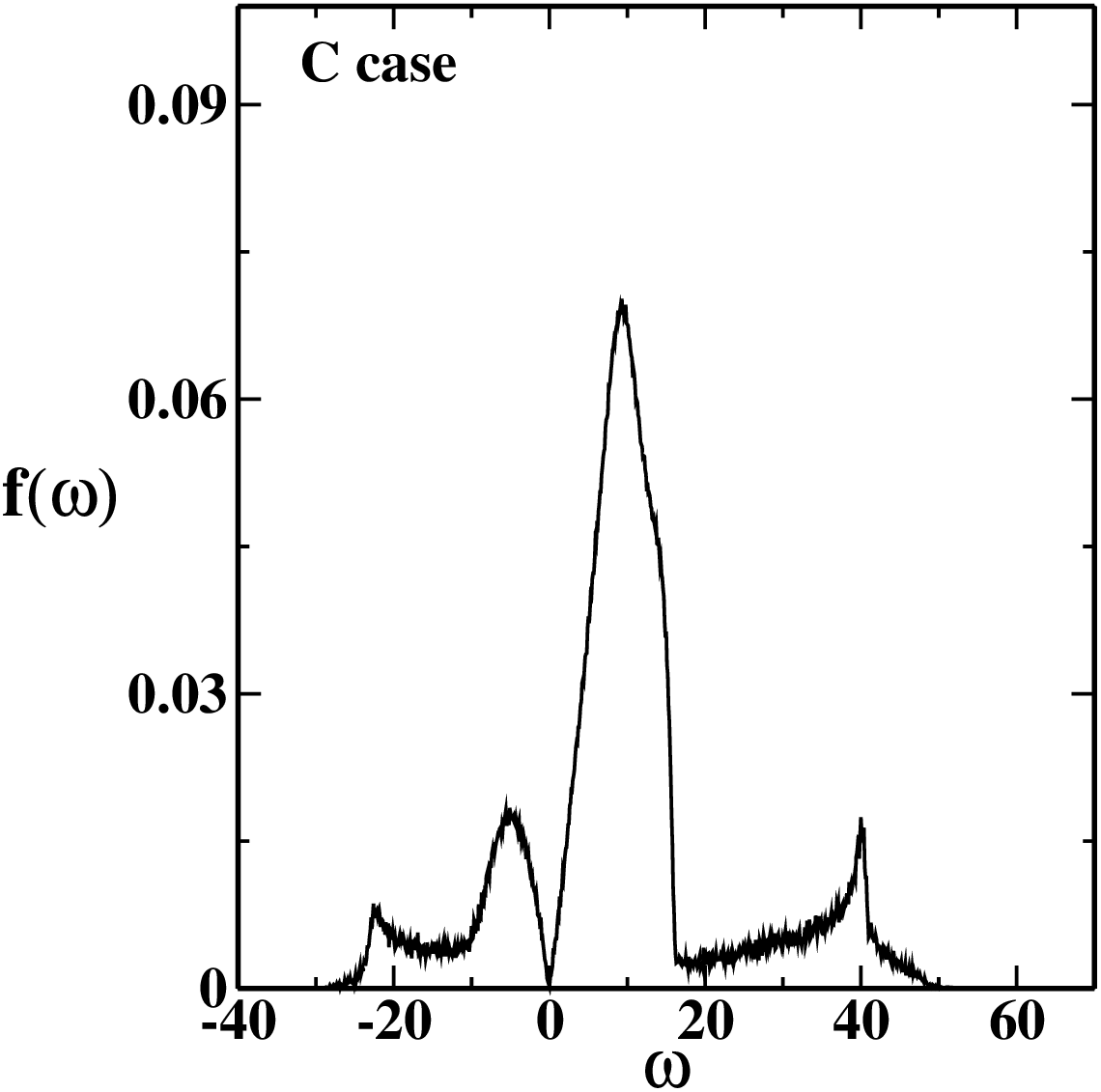}
   \includegraphics[clip=true,scale=0.35]{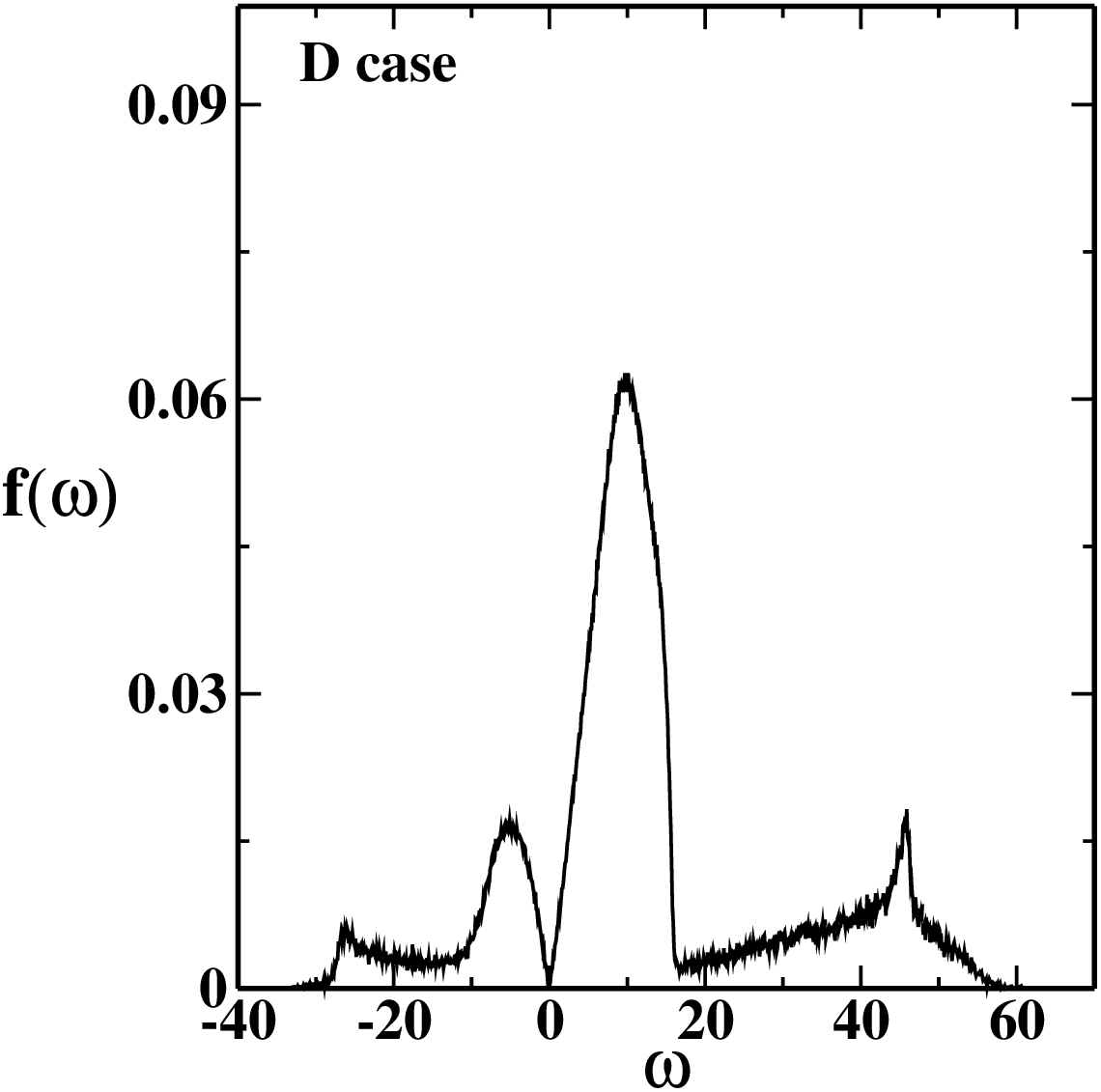}
   \caption{Normal models versus frequency for the  four studies cases. The density is fixed, $\rho^*=0.50$ in all the cases and the temperature is varied.} \label{fig:f-omega}
\end{figure}

The variation in anomalous behaviour in the multi-Gaussian  family of
water-like liquids studied here suggests that in addition to 
length scales, we need to look at other features of the pair potential 
e.g. its first and second derivatives. Instantaneous normal mode analysis
provides a way to summarize information on the curvature distribution of
the potential energy landscape. In Figure~\ref{fig:f-omega}, we show the INM spectra of
liquids bound by the four potentials (A, B, C and D)  at a common state 
point of $\rho^*=0.50$ and temperature $T^*=0.8$. The crucial 
features are as follows:
\begin{itemize}
\item A low-frequency split peak in the real branch centered at about $\omega =10$, that
does not vary significantly between the four cases and must reflect modes associated with the outer attractive well;
\item A high-frequency peak in the real branch, centered at approximately 30, 35, 40 and 50 for cases $A$, $B$, $C$ and $D$ respectively, which must correspond to 
motion in neighborhood of the shoulder length scale. As the curvature of the short-range minimum
increases, this features shifts to higher frequencies and becomes more prominent;
\item  The imaginary branch reflects regions of negative curvature in the neighborhood 
of barriers and inflexion points. Case $A$, where there is no barrier in the pair interaction but only an  inflexion point has a single peak like a 
simple liquid. However, this peak is broad because of the core-softened repulsive wall  and the
fraction of imaginary modes is large. As the barrier between the short and long length scales becomes more pronounced in the pair interaction, the 
second peak in the imaginary branch becomes more prominent. 
\end{itemize}

Thus the real branch is dominated by vibrational modes associated
with motion in the attractive and shoulder length scales while the  
imaginary modes branch is dominated by negative curvature modes
associated with transitions between the shoulder and attractive wells. For the 
case $A$, the real branch has three peaks related to the 
three basins: the shoulder scale, the attractive scale and a second
attractive scale located at further distance in Fig.~\ref{fig:potential}.
It has just one imaginary peak that indicating that
transitions between the two length scales do not require local barrier crossing.
For the cases $B$ and $C$ the imaginary branch has two peaks
suggestive of   modes connecting  between the shoulder 
scale, attractive scale and second attractive scales.
The peak with largest frequency in the real branch 
has larger frequency in the case $C$ than in the case $A$ and  is related to the shoulder scale.
For the case $D$, the shoulder is deep and so the  frequency related to the shoulder scale has a very
high frequency. The imaginary branch has two distinct oscillation modes that exclude transitions  between the 
shoulder scale and the other scales and  therefore no anomalies are expected.

\begin{figure}[htbp]
   \centering
   \includegraphics[clip=true,scale=0.32]{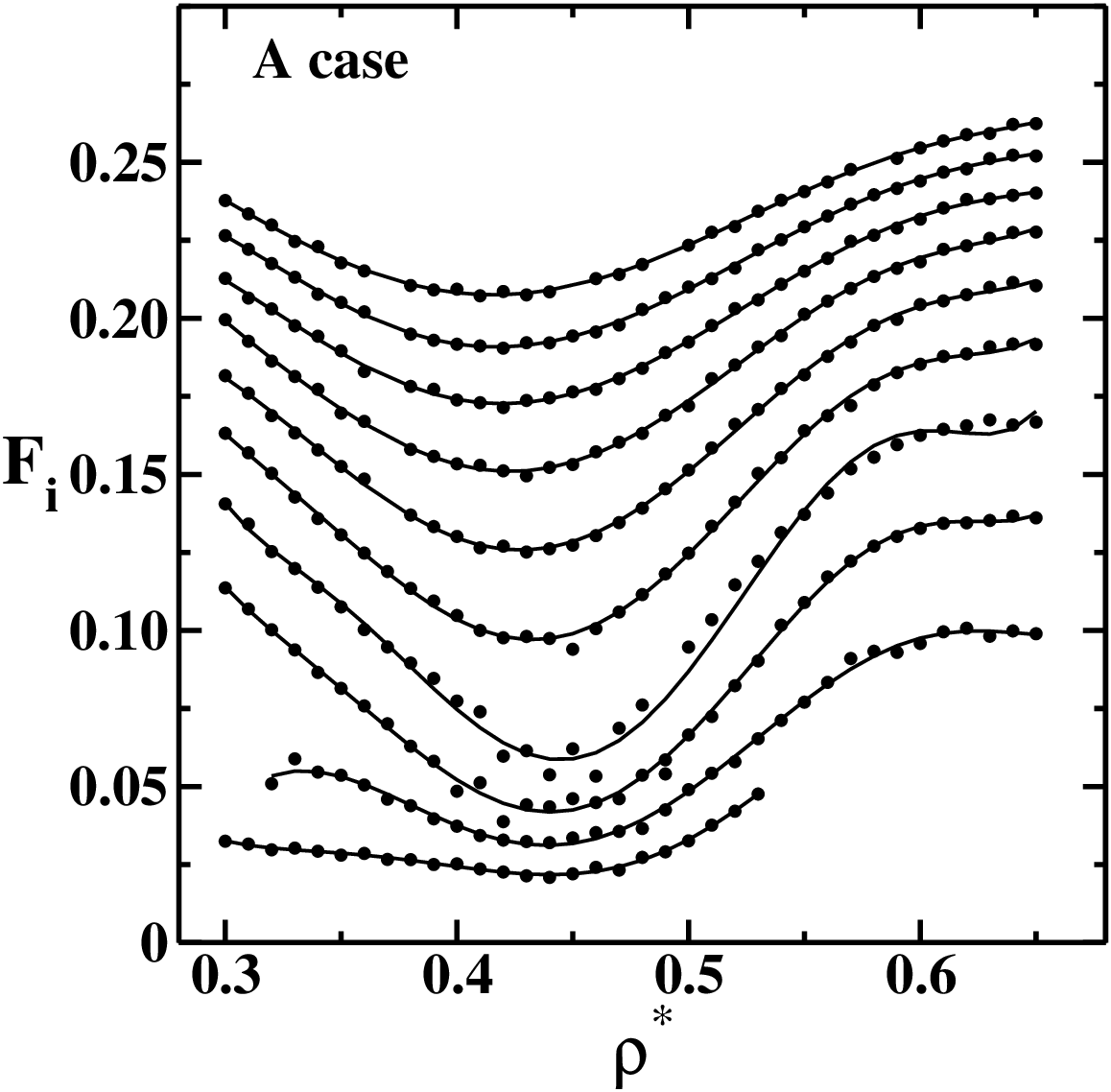} 
   \includegraphics[clip=true,scale=0.32]{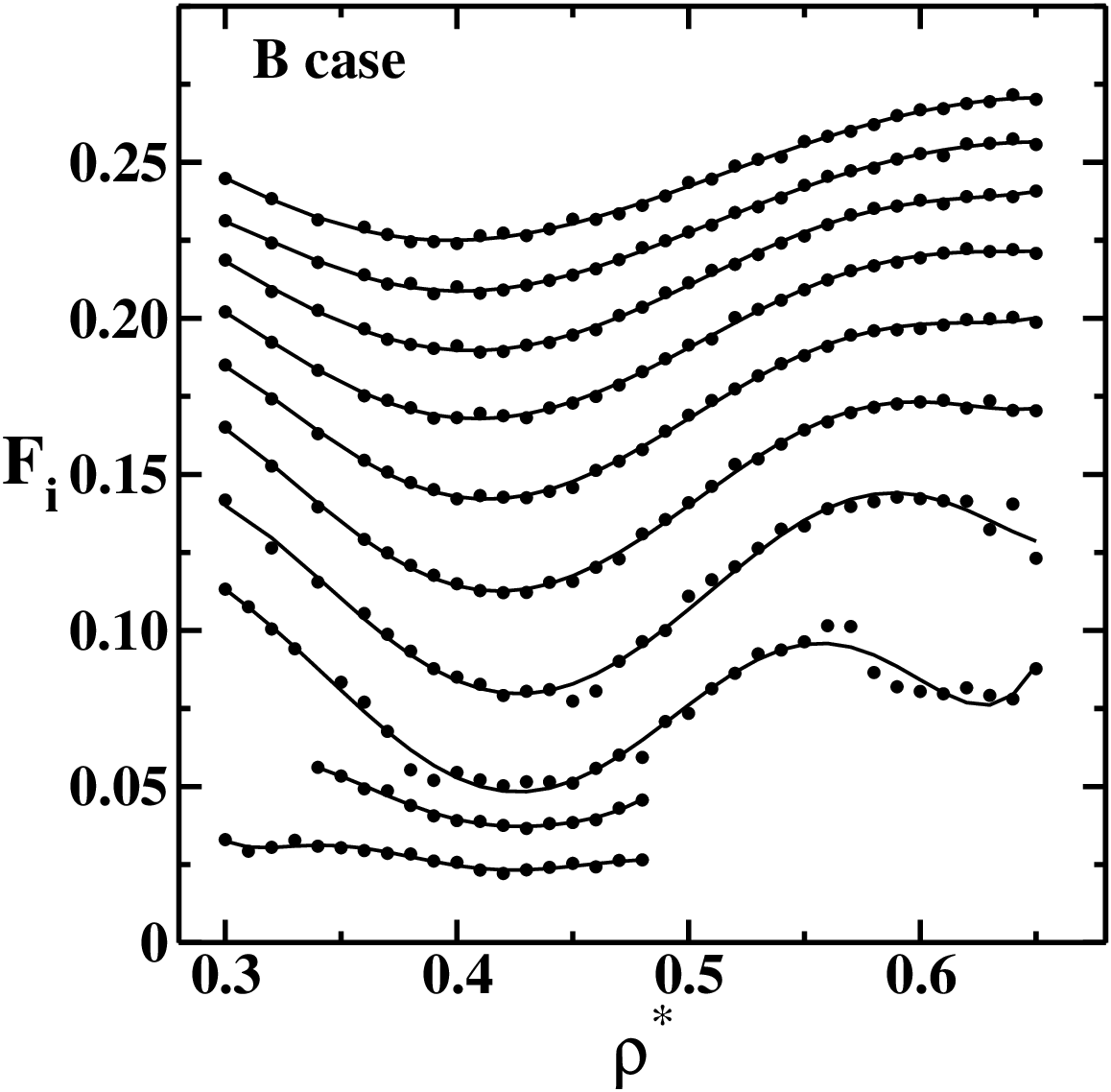} 
   \includegraphics[clip=true,scale=0.32]{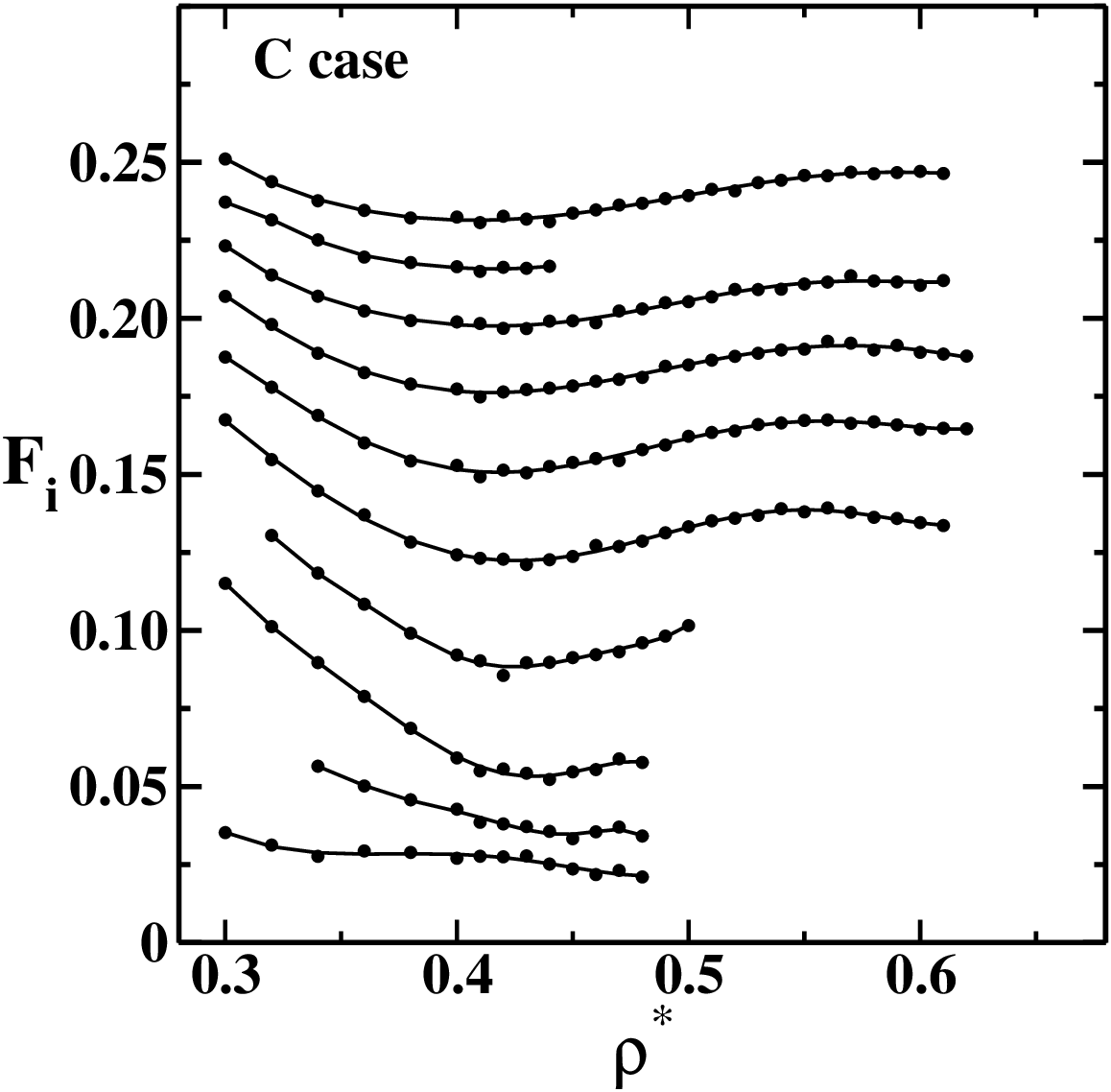} 
   \includegraphics[clip=true,scale=0.32]{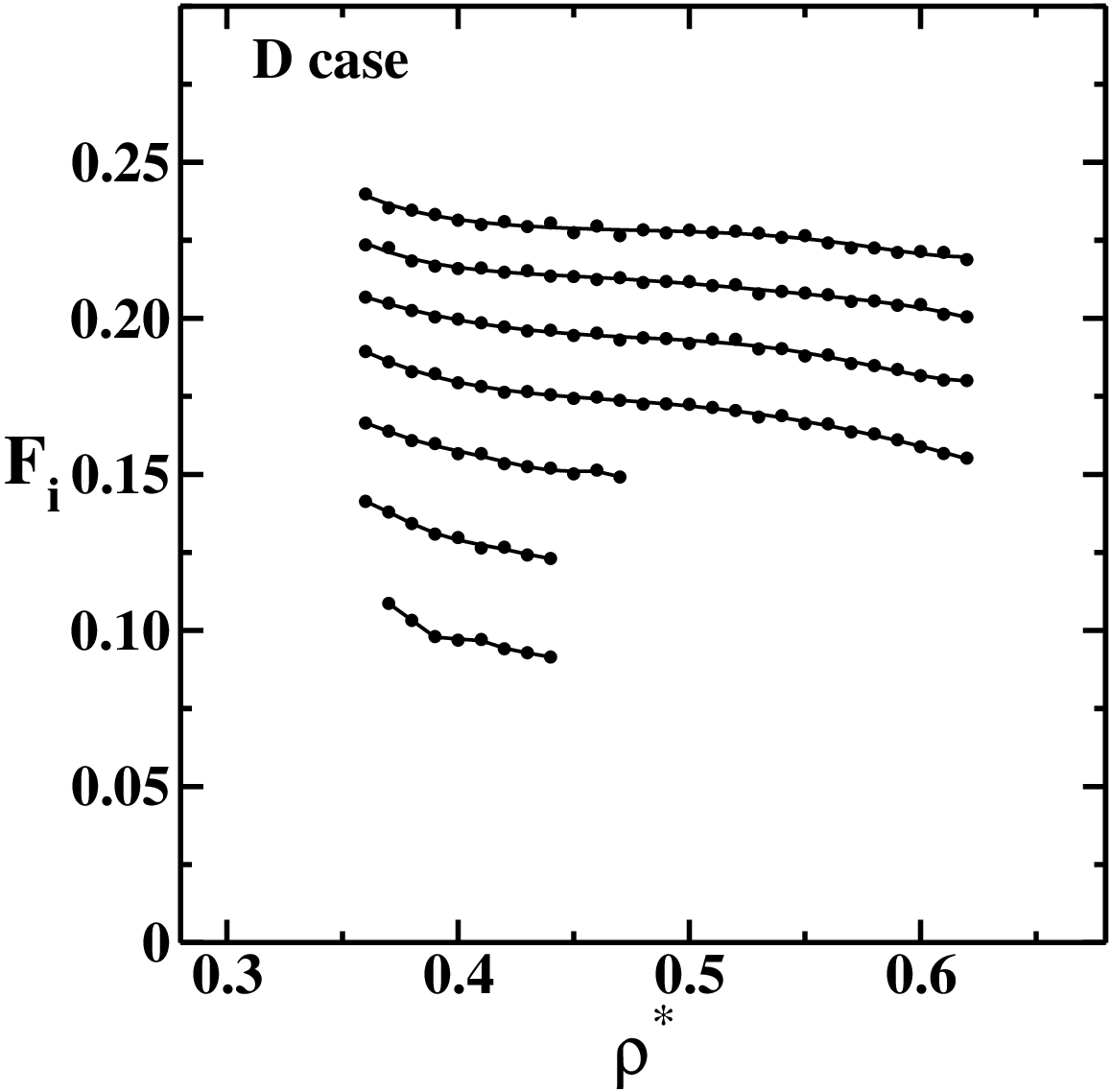} 
   \caption{Fraction of imaginary modes versus density for fixed temperatures, $T^* = 0.20, 0.30, 0.40, 0.50, 0.60, 0.70, 0.80, 0.90$ and $1.10 $ from bottom to
top for cases $A, B$ and $C$. For case $D$ the start at $T^*=0.50$.}
   \label{fig:Fi.vs.rho}
\end{figure}

The above discussion suggest that INM spectra carry fairly detailed information on the dynamics of transitions between the two length
scales. The two features which are a compact signature of INM spectra are the Einstein frequency and the fraction of imaginary modes.
Isotherms of the Einstein frequency as a function of density for all four cases show a monotonic increase with density and
do not show any significant signatures of the water-like anomalies. The fraction of imaginary modes, in contrast, correlates 
strongly with the anomalous behaviour of the pair entropy and the  diffusivity. Figure~\ref{fig:Fi.vs.rho}  shows the $F_i$ curves versus
density for various isotherms of all four multi-Gaussian model fluids  studied here. The parallel behaviour of the $s_2(\rho )$ and $F_i(\rho )$ curves at
corresponding isotherms is immediately obvious, though the $F_i(\rho)$ have a stronger non-monotonic behaviour
than $s_2(\rho)$ curves. This can be seen most clearly for a  high-temperature isotherm.

\section{Conclusions}

This paper examines the relationship between water-like anomalies and the
liquid-liquid critical point in a family of model fluids with
 multi-Gaussian,   core-softened pair interactions. The pair interaction
in this family of liquids is composed of a sum of Lennard-Jones and Gaussian
terms, in such a manner that the longer length scale associated with
a shallow, attractive well is kept constant while the shorter length scale
associated with the repulsive shoulder changes from an inflexion point to
a minimum of progressively increasing depth. The maximum depth of the shoulder
length scale is chosen so that the resulting potential reproduces the
oxygen-oxygen radial distribution function of the ST4 model of water.
As the energetic stabilization of the shoulder length scale increases, the 
liquid-liquid critical point shifts to higher temperatures and lower 
pressures. Simultaneously, the temperature for onset of the density
anomaly decreases and the region of liquid state anomalies in the
pressure-temperature plane diminishes. The condition for the presence
of anomalies is inconsistent with divergences near a critical point,
so that in the limiting case of maximum shoulder well depth, the anomalies
disappear.

To understand our results for the phase diagram and liquid-state anomalies
of the multi-Gaussian family of water-like fluids, it is important to
note that, in addition to the presence of two length scales, it is necessary
to consider the energetic and entropic effects as determined by local
minima and curvatures of the pair interaction. As the shoulder depth increases,
the pressure required to form the high density liquid decreases and the
temperature up to which the high-density liquid is stable increases. This explains
the shift of the liquid-liquid critical point to much lower pressures and
higher temperatures. To understand the entropic effects associated with the
changes in the interaction potential, we computed the pair correlation 
entropy and demonstrated the attenuation of the excess entropy anomaly as the shoulder
length scale changed from an inflexion point to a deep minimum. In conjunction
with Rosenfeld-scaling of transport  properties, this is consistent with 
the progressive loss of water-like thermodynamic, structural and transport anomalies.
The excess entropy anomaly in two-scale, isotropic fluids is due to a 
rise in entropy as a result of competition between two length scales at intermediate
densities. In the case of continuous potentials, the vibrational entropy 
associated with the two length scales becomes important. To index the overall
curvature distribution in the liquid, we have used instantaneous normal mode
analysis and shown the  fraction of imaginary frequency modes correlates 
well with the anomalous behaviour of the diffusivity and the pair correlation
entropy. A detailed analysis of the INM spectrum shows that as the shoulder
well increases in depth, there is a simultaneous rise in the positive curvature
associated with the shoulder minimum as well as the negative curvature of the
barrier separating the
shoulder minimum from the attractive minimum. Consequently,  the
vibrational entropy associated with pairs of particles separated
by the shoulder distance decreases, relative to that of pairs trapped
in the outer attractive well. Therefore the mixing entropy
due to the presence of two length scales is counteracted by the
changes in vibrational entropy associated by the two length scales.

A general conclusion that emerges from this study is that even though the ratio between the two length scales is important 
for locating the  temperature range of the anomalies~\cite{Ya07}, additional energetic and entropic effects associated with local
minima and curvatures of the pair interaction can play an important role. The liquid-liquid phase separation depends on the relative energies
associated with the two length scales whereas the water-like anomalies  depend upon a continuous rise in entropy as a function of isothermal
compression. A number of recent studies of core-softened fluids illustrate this conclusion. For example,  energetic and entropic effects play a very different
 role in the discrete and discontinuous versions of the shouldered well potential~\cite{Ol06a}.
In the discrete  case, the  enthalpic implications do not change significantly and the liquid-liquid critical point is not significantly different in the two systems. In contrast, the 
 continuous potential allows for a smooth transformation through a 
range of quasi-binary states from low- to high-density and shows water-like
anomalies. A more recent study of core-softened 
fluids~\cite{Si10} shows that  increasing the depth of the attractive well,
while leaving the shoulder feature constant, 
  results in disappearance of the anomalies while shifting the liquid-liquid
critical point to lower pressures and higher temperatures.

\bibliographystyle{aip}
\bibliography{allrefs,group,Biblioteca,books,B}
\subsection*{Acknowledgments}

This work is supported by the Indo-Brazil Cooperation Program
in Science and Technology of the CNPq (Brazil) and DST (India).This work 
is also
partially supported by the CNPq through the INCT-FCx.

\vspace{1cm}

\end{document}